# Optical Observations of Meteors Generating Infrasound – II: Weak Shock Theory and Validation

Elizabeth A. Silber[1*], Peter G. Brown[1] and Zbigniew Krzeminski[1]

[1] Department of Physics and Astronomy, University of Western Ontario, London, Ontario, Canada, N6A 3K7

*Corresponding author:

E-mail: esilber@uwo.ca

Mailing Address:

Department of Physics and Astronomy
1151 Richmond Street
University of Western Ontario
London ON
N6A 3K7
CANADA
Ph: +1-519-661-2111 x86393
Fax: +1-519-661-2033

## Abstract

We have recorded a dataset of 24 centimeter-sized meteoroids detected simultaneously by video and infrasound to critically examine the ReVelle [1974] weak shock meteor infrasound model. We find that the effect of gravity wave perturbations to the wind field and updated absorption coefficients in the linear regime on the initial value of the blast radius ($R_0$), which is the strongly non-linear zone of shock propagation near the body and corresponds to energy deposition per path length, is relatively small (<10%). Using optical photometry for ground-truth for energy deposition, we find that the ReVelle model accurately predicts blast radii from infrasound periods ($\tau$), but systematically under-predicts $R_0$ using pressure amplitude. If the weak-shock to linear propagation distortion distance is adjusted as part of the modelling process we are able to self-consistently fit a single blast radius value for amplitude and period. In this case, the distortion distance is always much less (usually just a few percent) than the value of 10% assumed in the ReVelle model. Our study shows that fragmentation is an important process even for centimeter-sized meteoroids, implying that $R_0$, while a good measure of energy deposition by the meteoroid, is not a reliable means of obtaining the meteoroid mass. We derived an empirical period-blast radius relation of the form $R_0 = 15.4\tau - 0.5$ ($\tau \leq 0.7$s) and $R_0 = 29.1\tau - 11.6$ ($\tau > 0.7$s) appropriate to cm-sized meteoroids. Our observations suggest that meteors having blast radii as small as 1m are detectable infrasonically at the ground, an order of magnitude smaller than previously considered.

# 1. Introduction

## 1.1 Meteor Generated Infrasound

Well documented and constrained observations of meteor generated infrasound [Edwards et al., 2008; Silber and Brown, 2014] are an indispensable prerequisite for testing, validating and improving theoretical hypersonic shock propagation and prediction models pertaining to meteors [e.g. ReVelle, 1974]. However, due to the lack of a sufficiently large and statistically meaningful observational dataset, linking the theory to observations had been a challenging task, leaving this major area in planetary science underexplored.

Infrasound is low frequency sound extending from below the range of human hearing of 20 Hz down to the natural oscillation frequency of the atmosphere (the Brunt-Väisälä frequency). Due to its negligible attenuation when compared to audible sound, infrasound can propagate over extremely long distances [Sutherland and Bass, 2004], making it an excellent tool for the detection and characterization of distant explosive sources in the atmosphere. Infrasound studies have gained momentum with the implementation of the global IMS network after the Comprehensive Nuclear Test Ban Treaty (CTBT) opened for signature in 1996. The IMS network includes 60 infrasound stations, 47 of which are presently certified and operational, designed with the goal of detecting a 1 kt (TNT equivalent; 1 kt = 4.185 x $10^{18}$ J) explosion anywhere on the globe [Christie and Campus, 2010].

Included among the large retinue of natural (e.g. volcanoes, earthquakes, aurora, lightning) [e.g. Bedard and Georges, 2000; Garces and Le Pichon, 2009] and anthropogenic (e.g. explosions, re-entry vehicles, supersonic aircraft) [Hedlin et al., 2002] sources of infrasound are meteors [ReVelle, 1976; Evers and Haak, 2001]. A number of meteoritic events have been detected and studied [e.g. Brown et al., 2008; Le Pichon et al., 2008; Arrowsmith et al., 2008] since the deployment of the IMS network. Often, no other instrumental records for these bolides are available; hence infrasound serves as the sole means of determining the bolide location and energy. A notable example of such an observation is the daylight bolide/airburst over Indonesia, which occurred on 8 October, 2009 and produced estimated tens of kilotons in energy [Silber et al., 2011].

Most recently, on 15 February, 2013, an exceptionally energetic bolide exploded over Chelyabinsk, Russia, causing significant damage on the ground as well as a number of injuries [Brown et al., 2013; Popova et al., 2013]. Such events attest to the need to better understand the nature of the shock wave produced by meteors.

The shocks produced by meteoroids may be detected as infrasound signals at the ground. As meteoroids enter the Earth's atmosphere at hypersonic velocities (11.2 – 72.8 km/s) [Ceplecha et al., 1998], corresponding to Mach numbers from ~35 to 270 [Boyd, 1998], they produce luminous phenomena known as a meteor through sputtering, ablation and in some cases fragmentation [Ceplecha et al., 1998]. Meteoroids can produce two distinct types of shock waves which differ principally in their acoustic radiation directionality. Their hypersonic passage through the atmosphere may produce a ballistic shock, which radiates as a cylindrical line source. Episodes of gross fragmentation, where a sudden release of energy occurs at a nearly fixed point [ReVelle, 1974; Bronshten, 1983] may result in an additional quasi-spherical shock [e.g. Brown et al, 2007; ReVelle, 2010]. Depending on a relative position of the meteor trajectory in respect to an observational point, as well as atmospheric propagation conditions, infrasonic signals from different types of shocks may be recorded [Silber and Brown, 2014].

Although infrasound does not suffer from significant attenuation over long distances, it is susceptible to dynamic changes that occur in the atmosphere. Nonlinear influences, atmospheric turbulence, gravity waves and winds, all have the potential to affect the infrasonic signal as it propagates between the source and the receiver [Ostashev, 2002; Kulichkov, 2004; Mutschlecner and Whittaker, 2010]. Consequently, distant explosive sources, such as bolides, are generally difficult to fully model or uniquely separate from other impulsive sources based on infrasound records alone.

The first complete quantitative model of meteor infrasound was developed by ReVelle [1974]. In this model predictions are made, starting with a set of source parameters, for the maximum infrasound signal amplitude and dominant period at the receiver. Due to a lack of observational data, ReVelle's [1974] cylindrical blast wave theory for meteors has never been experimentally and observationally validated. In particular, regional (<300 km) meteor infrasound signals have been studied infrequently in favor of larger bolide events, despite the fact that regional meteor infrasound is likely to reveal more characteristics of the source shock, having been substantially less modified during the comparatively short propagation distances involved [Silber and Brown, 2014].

A central goal of meteor infrasound measurements is to estimate the size of the relaxation or blast radius ($R_0$), as this is equivalent to an instantaneous estimate of energy deposition, which is the key to defining the energetics in meteoroid ablation. Indeed, all meteor measurements ultimately try to relate observational information back to energetics either through light, ionization or shock (infrasound) production. In order to better define meteoroid shock production, evaluate energy deposition mechanisms and estimate meteoroid mass and energy, it is helpful to first investigate near field meteor infrasound (ranges < 300 km) for well documented and characterized meteors, because this offers the most plausible route in validating the cylindrical blast wave model of meteor infrasound. Near field infrasonic signals are generally direct arrivals and suffer less from propagation effects.

In this work, we attempt to validate the existing ReVelle [1974] meteor infrasound theory, using a survey of centimeter-sized and larger meteoroids recorded by a multi-instrument meteor network [Silber and Brown, 2014; Silber, 2014]. This network, designed to optically detect meteors which are then used as a cue to search for associated infrasonic signals, utilizes multiple stations containing all sky video cameras for meteor detection and an infrasound array located near the geographical centre of the optical network.

## 1.2 Brief Review of ReVelle [1974] Meteor Weak Shock Theory

In the early 1950s, Whitham [1952] developed the F-function approach to sonic boom theory, a novel method of treating the flow pattern of shock signatures generated by supersonic projectiles, now widely used in supersonics and classical sonic boom theory [e.g. Maglieri and Plotkin, 1991]. It was soon realized that although the F-function offers an excellent correlation between experiment and theory for low Mach numbers (< ~3), it is not an optimal tool in the hypersonic regime [e.g. Carlson and Maglieri, 1972; Plotkin, 1989]. Recently, the Whitham F-function theory has been applied to meteor infrasound [Haynes and Millet, 2013], but it has not yet received a detailed observational validation. We note, however, that this approach offers another theoretical pathway to predicting and interpreting meteor infrasound, though we do not explore it further in this study.

Drawing on the early works of Lin [1954], Sakurai [1964], Few [1969], Jones et al. [1968], Plooster [1968; 1970] and Tsikulin [1970], ReVelle [1974; 1976] developed an analytic blast wave model of the nonlinear disturbance initiated by an explosive line source as an analog for a meteor shock.

In cylindrical line shock theory, the magnitude of the characteristic blast wave relaxation radius ($R_0$) is defined as the region of a strongly nonlinear shock.

$$R_0 = (E_0/p_0)^{1/2} \qquad (1)$$

Here, $E_0$ is the energy deposited by the meteoroid per unit trail length and $p_0$ is the ambient hydrostatic atmospheric pressure. Physically this is the distance from the line source at which the overpressure approaches the ambient atmospheric pressure. For a single body ablating in the atmosphere, ignoring fragmentation, the blast radius can be directly related to the drag force and ultimately expressed as a function of Mach number ($M$) and meteoroid diameter ($d_m$) [ReVelle, 1974]:

$$R_0 \sim M \, d_m \qquad (2)$$

While the original ReVelle [1974] model assumes propagation through an isothermal atmosphere, here we use an updated version incorporating a non-isothermal atmosphere. As shown in an earlier study [Edwards et al., 2008], the isothermal approximation leads to unrealistic values of signal overpressure. The following summary of ReVelle's [1974] meteor infrasound theory is similar to that presented in Edwards [2010], though with some corrections and emphasis on the approximations used by ReVelle [1974] and aspects of the treatment most applicable to our study.

A meteoroid generates the shock as it propagates through the atmosphere, but different parts of the shock (which have different blast radii at different points) will be recorded depending on the location of the station with respect to the orientation and geometry of the meteor trail. The ReVelle [1974] approach begins with a set of input parameters characterizing the entry conditions of the meteoroid, and from these initial conditions predicts the infrasonic signal overpressure (amplitude) and period at the ground. As part of this analysis, the blast radius and the height at which the shock transitions from the weakly nonlinear regime to the linear regime (distortion distance) are also determined. The model inputs are:

i.      station (observer) location (latitude, longitude and elevation);
ii.     meteoroid parameters (mass, density, velocity, and entry angle as measured from the horizontal);
iii.    infrasonic ray parameters at the source which reach the station based on ray-tracing results (angular deviation from the meteoroid plane of entry and shock source location along the trajectory in terms of latitude, longitude and altitude).

In the ReVelle [1974] meteor cylindrical blast wave theory the following assumptions are made:

i.      The energy release must be instantaneous.
ii.     The cylindrical line source is valid only if $v \gg c_s$ (the Mach angle has to be very small many meteoroid diameters behind the body) and $v$ = constant [Tsikulin, 1970]. Therefore, it follows that if there is significant deceleration ($v < 0.95v_{entry}$) and strong ablation, the above criteria are not met and the theory is invalid.
iii.    The line source is considered to be in the free field, independent of any reflections due to finite boundaries, such as topographical features [ReVelle, 1974].
iv.     Ballistic entry (no lifting forces present - only drag terms)
v.      The meteoroid is a spherically shaped single body and there is no fragmentation
vi.     The trajectory is a straight line (i.e. gravitational effects are negligible). The nonlinear blast wave theory does not include the gravity term.

The coordinate system to describe the motion and trajectory of the meteoroid, as originally developed by ReVelle [1974; 1976], is shown in Figure 1. Other treatments (e.g. Edwards [2010]) have not always correctly and consistently defined these quantities relative to the original definitions given in ReVelle [1974]. In this model, only those rays which propagate downward and are direct arrivals are considered (i.e. direct source-observer path). The predicted signal period, amplitude and overpressure ratio as a function of altitude for an example meteor are shown in Figure 2.

Note that due to severe nonlinear processes, the solutions to the shock equations are not valid for $x \leq 0.05$, where $x$ is the distance in units of blast radii (e.g. $R/R_0$). Once the wave reaches a state of weak nonlinearity (i.e. the shock front pressure ($p_s$) ~ ambient pressure at a given altitude ($p_0$)), the shock velocity approaches the local adiabatic speed of sound ($c$). When the relative overpressure $\Delta p$ (henceforward referred to as overpressure) is small, i.e. $\Delta p/p_0 \leq 1$ (at $x \geq 1$), weak shock propagation takes place and geometric acoustics becomes valid [Jones et al., 1968; ReVelle, 1974]. It is also assumed that at beginning, near the source ($x < 1$), the wave energy is conserved except for spreading losses [Sakurai, 1964].

Drawing upon theoretical and observational work on shock waves from lightning discharges [Jones et al., 1968] the functional form of the overpressure has limiting values of:

$$f(x) \atop x \to 0 = \frac{2(\gamma + 1)}{\gamma} \frac{\Delta p}{p_0} \to x^{-2} \quad (3a)$$

and

$$f(x) \atop x \to \infty = \left(\frac{3}{8}\right)^{-3/5} \left\{\left[1 + \left(\frac{8}{3}\right)^{8/5} x^2\right]^{3/8} - 1\right\}^{-1} \to x^{-3/4} \quad (3b)$$

Here, $\gamma$ is the specific heat ratio ($\gamma = C_p/C_v = 1.4$). In the limit as $x \to 0$, where $\Delta p/p_0 > 10$, attenuation is quite rapid ($x^{-2}$), transitioning to $x^{-3/4}$ as $x \to \infty$, where $\Delta p/p_0 < 0.04$ (or $M = 1.017$) [Jones et al., 1968]. Taking advantage of equations (3a) and (3b), and using results obtained from experiments [Jones et al., 1968; Tsikulin, 1970], the overpressure (for $x \geq 0.05$) can be expressed as:

$$\frac{\Delta p}{p_0} = \frac{2(\gamma + 1)}{\gamma} \left(\frac{3}{8}\right)^{-\frac{3}{5}} \left\{\left[1 + \left(\frac{8}{3}\right)^{\frac{8}{5}} x^2\right]^{\frac{3}{8}} - 1\right\}^{-1} \quad (4a)$$

The limit within which this expression is applicable is $0.04 \leq \Delta p/p_0 \leq 10$ [Jones et al., 1968]. The above expression can also be written as:

$$\frac{\Delta p}{p_0} \cong \frac{2\gamma}{\gamma + 1} \left[\frac{0.4503}{(1 + 4.803x^2)^{\frac{3}{8}} - 1}\right] \quad (4b)$$

After the shock wave has travelled a distance of approximately $10R_0$, where it is assumed that strong nonlinear effects are no longer important, its fundamental period ($\tau_0$) can be related to the initial blast radius via: $\tau_0 = 2.81R_0/c$, where $c$ is the local ambient thermodynamic speed of sound. The factor 2.81 at $x = 10$ was determined experimentally [Few, 1969] and found to compare favorably to numerical solutions [Plooster, 1968]. The frequency of the wave at maximum is referred to as the 'dominant' frequency [ReVelle, 1974]. For a sufficiently large $R$, and assuming weakly nonlinear propagation, the line source wave period ($\tau$) for $x \geq 10$ is predicted to increase with range as:

$$\tau(x) = 0.562 \, \tau_0 \, x^{1/4} \quad (5)$$

Far from the source, the shape of the wave at any point will mainly depend on the two competing processes acting on the propagating wave: dispersion, which reduces the overpressure and 'stretches' the period; and steepening, which is the cumulative effect of small disturbances, tending to increase the overpressure amplitude [ReVelle, 1974]. In ReVelle's [1974] model, however, it is assumed that the approximate wave shape is known at any point. After a short distance beyond $x = 10$, the waveform is assumed to remain an N-wave [DuMond et al., 1946] type-shape [ReVelle, 1974].

For the analytic implementation of ReVelle's theory, it is necessary to choose some transition distance from the source where we consider the shock as having moved from weakly nonlinear propagation to fully linear. The precise distance at which the transition between the weak shock and linear regime occurs is poorly defined. Physically, it occurs smoothly, as no finite amplitude wave propagating in the atmosphere is truly linear; this is always an approximation with different amplitudes along the shock travelling with slightly different speeds. This distance was originally introduced by Cotten et al. [1971] in the context of examining acoustic signals from Apollo rockets at orbital altitudes. Termed by Cotten et al. [1971] the "distortion distance", it is based upon Towne's [1967] definition of the distance ($d'$) required for a sinusoidal waveform to distort by 10%. ReVelle [1974] adopted this distance, together with the definition of Morse and Ingard [1968], to define the distance ($d_s$) an initially sinusoidal wave must travel before becoming "shocked". Thus, it follows that $d_s = 6.38\ d'$, where $d' > d_a$ and $d_a$ is the remaining propagation distance of the disturbance before it reaches the observer. Further details summarizing the main points of the ReVelle [1974] model germane to our study are given in the supplemental material.

In summary, according to the ReVelle [1974] weak shock model, there are two key sets of expressions to estimate the predicted infrasonic signal period and the amplitude at the ground. The first is the expression for the predicted dominant signal period in the weak shock regime ($d' \leq d_a$). Once the shock is assumed to propagate linearly, by definition the period remains fixed.

The second expression relates to the overpressure amplitude. In the weak shock regime the predicted maximum signal amplitude is given by:

$$\Delta p_{z \rightarrow obs} = [\ f(x)\ D_{ws}(z)\ N^*(z)\ Z^*(z)\ ]\ p_0 \qquad (6a)$$

where $f(x)$ is the expression given in equation (4b), $N^*$ and $Z^*$ are the correction factors as described in supplemental material and $D_{ws}$ is the weak shock damping coefficient.

Once the wave transitions into a linear wave, the maximum signal amplitude is given by:

$$\Delta p_{z \rightarrow obs} = \left[ \Delta p_{z \rightarrow t}\ D_l(z) \frac{N^*(z)_{z \rightarrow obs}\ Z^*(z)_{z \rightarrow obs}}{N^*(z)_{z \rightarrow t}\ Z^*(z)_{z \rightarrow t}} \left( \frac{x_{z \rightarrow t}}{x_{z \rightarrow obs}} \right)^{1/2} \right] p_0 \qquad (6b)$$

where $\Delta p_{z \rightarrow t}$ is identical to the expression given in equation (6a) and $D_l$ is the linear damping coefficient. The subscripts $z$, $t$ and $obs$ in equations (6a) and (6b) denote the source altitude, transition altitude and the receiver's altitude, respectively, following the notation in ReVelle [1974].

The ReVelle [1974] model as just described has been coded in *MATLAB*® to allow comparison between the predicted amplitudes and periods of meteor infrasound at the ground with the observations, the focus of this paper. In our first paper in the series [Silber and Brown, 2014], we used optical measurements to positively identify infrasound from 71 meteors and constrain the point (and its uncertainty) along the meteor trail where the observed infrasound signal emanated. Some of these meteors produced multiple infrasound signals (two or three) as recorded at the station, bringing the total number of distinct arrivals to 90. We also developed a meteor infrasound taxonomy using the pressure-time waveforms of all the identified meteor events as a starting point to gain insight into the dominant processes which modify the meteor infrasound signal as observed at the ground. Furthermore, that work examined the influence of atmospheric variability on near-field meteor infrasound propagation and established the type of meteor shock production at the source (spherical vs. cylindrical). Nearly 70% of arrivals were most likely produced by the cylindrical line source type of shock.

Here, we use the dataset constructed in the first part of our study and select the best constrained (i.e. those for which we have accurate infrasound source heights from raytrace solutions and high fidelity optical measurements solutions and are most likely cylindrical and not spherical shocks) meteor events to address the following:

i.   for meteors detected optically and with infrasound, use the ReVelle [1974] weak shock theory to provide a bottom-up estimate of the blast radius (i.e. from observed amplitude and period at the ground can we self-consistently estimate the blast radius at the source);

ii.  test the influence of atmospheric variability, winds, Doppler shift and initial shock amplitude on the weak shock solutions within the context of ReVelle [1974] meteor infrasound theory;

iii. determine an independent estimate of meteoroid mass/energy from infrasonic signals alone and compare to photometric mass/energy measurements;

iv.  critically evaluate and compare ReVelle's [1974] weak shock theory with observations, establishing which parameters/approximations in the theory are valid and which may require modification.

## 2. Methodology and Results

### 2.1 Weak Shock: Model Updates and Sensitivities

The ReVelle [1974] weak shock model algorithm was updated to include the full wind dependency and Doppler shift for the period [Morse and Ingard, 1968] as a function of altitude. We performed a sensitivity study to test the effects of: (i) absorption coefficients (supplemental material) in the linear regime using the set given by ReVelle [1974] and that of Sutherland and Bass [2004]; and (ii) gravity wave perturbations to the wind field [Silber and Brown, 2014]. A brief methodology outlining the model updates and the sensitivity study are given in the supplemental material. The overall effect of both winds and Doppler shift on the weak shock model was found to be relatively small, resulting in $R_0$ differences of no more than 13% for the period (average 4%) and as high as 9% for the amplitude (average 3%). The perturbations to the atmospheric winds expected from gravity-waves were found to have even smaller effects on estimates of $R_0$, typically of 10% or less.

In addition, the predicted ground-level period and amplitude outputs of the weak shock model were tested using a synthetically generated meteor (Figure 3).

## 2.2 Weak Shock: Bottom-up Modelling

The first approach we adopt to testing the ReVelle [1974] theory is a bottom-up methodology. This provides an indirect method of estimating the blast radius at the meteor using infrasound and optical astrometric measurements of the meteor only (i.e. without prior independent knowledge of the meteoroid mass and density). Given the input parameters, all of which are known except the blast radius, the goal is to answer the following question: what is the magnitude of the blast radius required to produce the observed signal amplitude and period at the station if we assume (i) the signal remained a weak shock all the way to the ground and (ii) if it transitioned to the linear regime? Additionally, we want to define the blast radius uncertainty given the errors in signal measurements.

Drawing upon the results obtained in Silber and Brown [2014], the 24 well constrained optical meteors which were also consistent with cylindrical line sources (as determined through optical measurements and raytracing) were used to observationally test the weak shock model. Out of these 24 events, 18 produced a single infrasonic arrival, while six events produced two distinct infrasonic arrivals at the station. All 24 events in this data set satisfy the condition of negligible deceleration prior to the onset of shock. The orbital parameters and meteor shower associations for our data set are listed in Table S1. The summary of event parameters is given in Table S2. The meteor shock source altitude in our data set ranges from 53 km to 103 km, the observed signal amplitude ($A_{obs}$) is from 0.01 Pa to 0.50 Pa, while the observed dominant signal period ($\tau_{obs}$) is between 0.1 s and 2.2 s. Typical values of overpressure from meteors in this study are 1-2 orders of magnitude smaller than those associated with the signals from Apollo rockets as reported by Cotten et al. [1971], the last comparable study to this one.

For the model to be self-consistent a single blast radius should result from the period and amplitude measurements. In practice, we find estimates of blast radii for period and amplitude independently in both linear and weak shock regimes such that the measured signal amplitude or period is matched within its measurement uncertainty. Therefore, for each amplitude/period measurement there are two pairs of theoretical quantities produced from ReVelle's [1974] theory: the predicted signal amplitude ($A_{ws}$) and period ($\tau_{ws}$) in the weak shock regime, and the signal amplitude ($A_l$) and period ($\tau_l$) in the linear regime. The iterations began with a seed-value of the initial $R_0$, and then based on the computed results the process is repeated with a new (higher or lower) value of $R_0$ and the results again compared to measurements until convergence is reached. The result of this bottom-up procedure is a global estimate of the blast radius matching the observed amplitude or period assuming either weak-shock or linear propagation ($A_{ws}$, $\tau_{ws}$, $A_l$, $\tau_l$). In the second phase of this bottom-up approach, this global modelled initial blast radius was used as an input to iteratively determine the minimum and maximum value of the model $R_0$ required to match the observed signal (period or amplitude) within the full range of measurement uncertainty.

The summary of the bottom-up blast radius modelling results are presented in Figure 4. There is a significant discrepancy between the period-based blast radii and amplitude-based blast radii in both the linear and weak shock regimes. The average blast radius from amplitude determinations in the linear regime is approximately 30 times smaller than that in the weak shock regime, indicating that the transition to linearity approach employed in the ReVelle [1974] weak shock model perhaps significantly underestimates the blast radius. In the weak shock regime, where we assume the signal can be treated as a weak shock all the way to the ground, we find that the amplitude-estimated $R_0$ in most cases is larger than that estimated from the period, but the difference is much smaller than the linear case, being no more than a factor of 15 between the amplitude and period. Almost half of the events show agreement within uncertainty.

## 2.3 Top-Down Weak Shock Modelling

We use the fragmentation (FM) model of meteoroid motion, mass loss and radiation in the atmosphere [Ceplecha and ReVelle, 2005] to fit the observed brightness and length vs. time measured for the infrasound-producing meteor. As fragmentation is explicitly accounted for, the FM model should provide a more realistic estimate of energy deposition along the trail, and thus $R_0$.

The first step in constructing an entry model solution was to begin with the approximate (starting) values for intrinsic shape density coefficient ($K$) and intrinsic ablation coefficient ($\sigma$), which we then modified together with the initial mass in a forward modeling process. The observed light curve from photometric measurements (supplemental material), in conjunction with the astrometric solution (event time, path length as a function of height and entry angle) for each event, was used to match the theoretical light curve and model length vs. time to observations through forward modelling (Figure S1).

Depending on the shape of the light curve (e.g. obvious flares or a smooth light curve), the FM model was used to implement either a single body approach or discrete fragmentation points. While the FM model performs very well in matching the observed light curves, it should be noted that the final solution is representative, but not necessarily unique. For example, if the shape coefficient is increased, then similar output results are found by reducing the initial mass. These differences in initial parameters and their variation at fragmentation points may result in an uncertainty of up to factor of several in the initial mass and therefore affect the mass loss in a similar fashion as a function of altitude. The ranges of the $\sigma$ and $K$ values used for modelling are given in Table S3. Further details about the FM model, including input and output parameters, are described in the supplemental material.

In the general case, the energy lost by the meteoroid per unit path length is:

$$dE/dL = (v^2/2)\,(dm/dL) + mv\,(dv/dL) \tag{7}$$

To determine the blast radius using the dynamics from the FM model, we applied equation (7), also accounting for mass loss during fragmentation episodes, or single-body mass at the height corresponding to the shock height. We also included the altitude region corresponding to the uncertainty in the source height. Essentially, the blast radius derived from the FM model is an average of energy deposition (equation (1)) along the source height ± height uncertainty.

However, since the height uncertainties are very small, the uncertainty in the blast radius is negligible (the error bars are shown in Figure 5).

To perform the top-down weak shock modelling, we start at the "top" by using the FM model estimates for $R_0$ in conjunction with the other optically measured parameters for each meteor and calculate the predicted signal period and amplitude at the ground, assuming both weak shock and linear propagation. This provides an independent comparison between the bottom-up modeling and forms a potential cross-calibration between infrasonically derived energy/mass and the same photometrically estimated quantities.

Figure 5 shows the comparison of the blast radius as obtained via FM model versus the blast radius from bottom-up modelling including all four outputs (the period and amplitude in linear and weak shock regimes). In the linear regime, the amplitude-based $R_0$ from the bottom-up modelling is clearly underestimated compared to the $R_0$ derived from the FM model. The comparison between the predicted signal period and amplitude using $R_0$ from the top-down modelling and that observed is shown in Figure 6. The predicted signal amplitude is somewhat underestimated in the weak shock regime and markedly overestimated in the linear regime. The predicted signal period, however, shows a near 1:1 agreement with the observed period.

### 2.4 Infrasonic Mass

From the optical measurements of the meteor and appealing to the fireball classification work of Ceplecha and McCrosky [1976] it is possible to correlate the fireball type with its likely physical properties (i.e. density) [Ceplecha et al., 1998] (Table S4) and thus compute an infrasonic mass [Edwards et al., 2008] through $m = \pi \rho d_m^3/6$, assuming a spherical shape and single body ablation (i.e. no fragmentation) from the bottom-up modelling. Using equation (2) and the relationship between the meteoroid density and diameter, infrasonic mass ($m_{infra}$) is then given by:

$$m_{infra} = (\pi \rho/6)(R_0/M)^3 \qquad (8)$$

Considering that the bottom-up modelling yields four values of blast radii as described in section 2.2, there are four resultant infrasonic masses.

The meteoroid mass can also be estimated through several empirical meteor magnitude-mass-speed relations as determined from earlier photographic surveys [Jacchia et al., 1967], or by summing the total light emission [Ceplecha et al., 1998; Weryk and Brown, 2013] and assuming the luminosity is a known fraction of the total kinetic energy loss. The meteor mass estimated through light production (called photometric mass) can be independently checked against the mass estimated from the observed deceleration of the meteor (termed dynamic mass). These approaches and the detailed results of camera estimates of meteoroid energy converted to equivalent blast radius are described in the supplemental material. The infrasonic masses derived from the period-based $R_0$ are in better agreement with the photometric masses than are the infrasonic masses derived from the amplitude-based $R_0$ in either regime (Figure S2). Additionally, drawing upon equation (1), meteoroid energy can be derived from $R_0$ (bottom-up and top-down) without needing to make any assumption about single body (Figure S3).

## 3. Discussion

ReVelle [1974] suggested that the lower threshold for the blast radius of meteors generating infrasound, which should be detectable at the ground, should be in the range of ~5 m. Having well constrained observational data, we derive the blast radii between 1.1 m and 51 m from our best fit bottom-up weak shock modelling and 0.4 – 41 m from the top-down FM model. A smaller blast radius is typically associated with lower source altitudes (< 80 km). This suggests that the original ReVelle [1974] estimate may well be too high by a factor of almost ten.

The dominant signal period is more robust than signal amplitude when estimating meteoroid energy deposition, as it is less susceptible to adverse propagation effects in the atmosphere [ReVelle, 1974; Edwards et al., 2008; Ens et al., 2012]. The dominant signal period is proportional to the blast radius, and therefore energy deposition by a meteoroid (mass and velocity) and the shock altitude. We remark that the signal period undergoes very small overall changes during propagation as it changes slowly in the weak shock regime (e.g. equation (5), Figure 2) and remains constant once it transitions to the linear regime. Therefore, the weak shock period is closer to the fundamental period at the source and expected to be a more robust indicator of the initial blast radius and hence energy of the event.

In contrast, the signal amplitude is generally more susceptible to a myriad of changes during propagation. The effects of non-linearity and wave steepening, as well as the assumed transition point from weak shock to a linear regime of propagation, are poorly constrained for meteors. Indeed, accurately predicting signal amplitudes for high altitude infrasonic sources (e.g. meteors), especially in the linear regime has been recognized as a long standing problem [e.g. ReVelle, 1974; 1976; Edwards et al., 2008]. Edwards et al. [2008] noted the significant differences in predicted amplitude for simultaneously observed optical-infrasound meteors, especially that in the linear regime, and determined that meteor infrasound reaches the ground predominately as a weak shock. While a similar empirical deduction can be made from Figure 4 in our study, this does not physically explain the discrepancy.

As shown earlier through top-down modeling, we expect the period-estimated blast radius in either regime to be relatively close to true values, as the observed period is much less modified during propagation compared to the amplitude. With this in mind, it becomes evident that the transition altitude is a major controlling factor in the linear amplitude predictions. We observe that the consistently smaller amplitude-based blast radius in the linear regime originates from the fact that depending on the transition altitude, the amplitude grows in the linear regime as a result of the change in ambient pressure. This leads to two questions: (i) is it possible to find the distortion distance and therefore a transition height which would predict both the period and amplitude such that the observed quantities are matched? (ii) given enough adjustment in the distortion distance, would any wave eventually transition to the linear regime?

We investigated the effect of the transition altitude by varying the distortion distance. The distortion distance was originally defined as the distance a wave would have to travel before distorting by 10% [Towne, 1967]. The original definition of the distortion distance [Towne, 1967] is:

$$d' = \lambda \; [20 \; (\gamma + 1) \; S_m]^{-1} = c\tau \; [34.3 \; (\Delta p/p)]^{-1} \qquad (9)$$

where $\lambda$ is the wavelength, $S_m$ is the overdensity ratio ($\Delta\rho/\rho_0$), $c$ is the adiabatic speed of sound and $\Delta p/p$ is the overpressure. The distortion distance is a constant number of wavelengths for waves of different frequency [Towne, 1967]. There are two major assumptions in equation (9): (i) the wave is initially sinusoidal, (ii) $S_m$ is small but not negligible. An intense sound wave ($S_m = 10^{-4}$) would therefore distort by 10% within 200 wavelengths [Towne, 1967; Cotten et al., 1971].

By varying the constant in the denominator in the right hand side of equation (9) to reflect an 'adjustable' distortion distance factor and by using the weak shock period-determined $R_0$ value as an input, a series of bottom-up modelling runs were performed. These were aimed at finding simultaneously both the predicted linear regime amplitude and the period which would match the observed quantities within their measurement uncertainty (Figure 6). Physically, this corresponds to adopting a series of different distortion distance definitions (rather than using the original distance with 10% distortion remaining assumption) with the goal of matching the signal properties and deriving the new set of bottom-up $R_0$. Henceforth, these new blast radii are referred to as the best fit $R_0$. The outcome of this investigation was:

- It was possible to find the converging amplitude-period solution in the linear regime for the majority of arrivals (22/24). Three fits are as much as 20% beyond the measurement uncertainty bounds for amplitude, while all others are within the measurement uncertainty bounds for both period and amplitude.

- When varying the constant in equation (9), the resultant distortion distance percentage is well below 6%. The distribution is shown in Figure 7d. Moreover, it is not possible to define a set percentage that would be applicable to all events. Some of these values may in fact not be realistic or feasible - we are assuming that the entire cause of the difference in the linear amplitude vs. period is due to the definition of the distortion distance, an assumption probably not fully correct.

- Smaller distortion distance leads to lower transition altitudes (Table S5). Half of the arrivals (12/24) had their transition height below 5 km. The maximum transition altitude was 25 km, with the mean altitude of 9 km; this means that no weak shock wave can be approximated as transitioning to a linear acoustic wave prior to reaching this height within the context of the original ReVelle [1974] model and still produce physically reasonable amplitude estimates. With the original definition for $d'$, the transition altitudes for our data were as high as 56 km, with a mean of 33 km. Cotten et al. [1971] examined propagation of shock waves generated by Apollo rockets and noted that a wave would not be expected to be acoustic above the altitude of 35-40 km at the large blast radii characteristic of those vehicles.

- Regardless of the extent of adjustment to the distortion distance, within the context of the ReVelle [1974] model, self-consistent amplitudes and periods are not possible unless some weak shock waves are assumed to never transition to linear waves. – i.e. even using a 0% distortion in our data set, two arrivals had no transition at all. However, these two arrivals also had a poor fit overall for any distortion distance.

- The lower transition altitude also implies that the difference between the classical and Sutherland and Bass [2004] absorption coefficients is negligible in the frequency range of our events.

We compared the infrasonic masses from the best fit $R_0$ to the photometric masses (Figure S6(a)) and found a much better agreement (on average to within order of magnitude) than that resulting from the comparison between photometric and the infrasonic masses from $R_0$ using the original definition of distortion distance.

The mass from the FM model $R_0$ should not exceed the mass used as the model input; however, the contrary can be observed (Figure S4(b)). For a single spherical body and no fragmentation and/or significant ablation, the blast radius can be estimated via equation (2). However, if there is fragmentation and/or significant ablation, then the contributions from the particles/fragments falling off the main body may alter the blast radius (Figure S5) such that there is an over-prediction of the meteoroid mass. Therefore, in such instances, the blast radius, while a good measure of energy deposition by the meteoroid ($R_0 \sim dE/dL$), is not a reliable means of obtaining the meteoroid mass. Recall from section 1.2 that there are several crucial assumptions in the weak shock model: (i) the meteoroid is a spherically shaped single body, (ii) there is no fragmentation, and (iii) there is no significant deceleration and strong ablation. ReVelle [2010] suggested that $R_0$ from a fragmenting bolide may be as much as 5-20 times larger (depending on height) than a non-fragmenting blast radius. Our study shows that strong ablation is indeed an important effect, even for centimeter sized meteoroids. In fact, this contribution is up to a factor of 10 and on average a factor of 3. Therefore, use of the $R_0 = Md_m$ is not valid in most cases. The blast radius estimated from purely energy-based (optical) considerations appears to be more robust.

Having done bottom-up and top down analysis, and blast radius and mass comparisons, it is evident that without having some information about the source function (i.e. measurements from video observations) the signal amplitude exhibits too much scatter to be utilized in empirical blast radius estimates. Although the dominant signal period may undergo variations due to competing processes of distortion and dispersion [ReVelle, 1974], this study demonstrates that the dominant signal period is much closer to 'true' values of $R_0$ (e.g. Figures 5 and 7(a)).

In Figure 8 we show best fit $R_0$ derived through the bottom-up modelling. For the centimeter-sized meteoroids we may estimate an empirical period-$R_0$ relation for the blast radius from these bottom-up values. Based on the best fits, there seem to be two distinct blast radius prediction populations: the short period ($\tau_{obs} \leq 0.7$s) and the long period ($\tau_{obs} > 0.7$s). The short period population, predicting $R_0 \leq 10$ m, is confined to the shock source altitudes between 53 km and 95 km. The long period population, predicting $R_0 > 10$ m, is associated with the shock source altitudes extending above 85 km. The residuals in the fits imply that for the short period population, the bottom-up based blast radius can be estimated within 1.5 meters, and for long periods within ~5 m, although the number statistics for the latter is small. We suspect much of the uncertainty at larger periods reflects the greater role fragmentation may play for larger meteoroids. The overlap altitude (85 − 95 km) between these two populations occurs at the mesopause and near the transitional slip-flow regime for meteoroids ~1 centimeter in diameter [Campbell-Brown and Koschny, 2004], suggesting that the weak shock model may not be correctly predicting $R_0$ in the free molecular flow regime. Further studies with larger number statistics are recommended to examine this aspect of the weak shock theory [ReVelle, 1974], as well as investigate a possible influence of instrumental sensitivity and noise levels at a recording station on the long/short period population distribution.

For completeness, we also fitted the $R_0$ calculated from the FM model. Recall that the FM based $R_0$ is derived from the energy deposition determined by fitting optical measurements to the entry model. The FM based $R_0$ has much more scatter, thus resulting in the blast radius estimate uncertainty of up to 17 m.

The empirical relations for the blast radius derived from the bottom-up approach are:

$$R_0 = 15.4\,\tau - 0.5 \quad (\tau_{obs} \leq 0.7\text{s}) \tag{10a}$$

$$R_0 = 29.1\,\tau - 11.6 \quad (\tau_{obs} > 0.7\text{s}) \tag{10b}$$

Even though there are a number of simplifications and assumptions in the weak shock model, in its new modified form it offers a reasonable initial estimate for the blast radius as a function of observed signal period, and therefore energy deposition for small regional meteoroids, without making any assumptions about the fragmentation process. Consequently, this methodology could be extended to high altitude explosive sources in the atmosphere. The applicability of the weak shock model can also be further investigated and extended to spherical shocks, especially if observational data can be used in a similar fashion as in this study.

## 4. Conclusions

In this paper we extend the study of Silber and Brown [2014] to critically investigate the analytic blast wave model of the nonlinear disturbance initiated by an explosive line source as an analog for a meteor shock as developed by ReVelle [1974]. We applied the updated ReVelle [1974] model to 24 of the best constrained simultaneously optical and infrasound detected regional meteors to critically examine the weak shock model. Here we summarize our main conclusions:

i.   We analyzed the weak shock model behavior using a synthetic meteor, and performed the sensitivity analysis on a set of meteor events from our data set to examine the effect of the winds, perturbed wind fields due to gravity waves, and Doppler shift. The overall effect of these factors on the initial value of the blast radius is relatively small (<10%) for regional meteor events.

ii. We performed bottom-up modelling using the ReVelle [1974; 1976] approach to determine the blast radius required to predict both the signal amplitude and period at the ground such that it matches that observed by the receiver within uncertainty. While the period-based $R_0$ appears to have realistic values, the amplitude-based $R_0$ exhibits large systematic deviations in the linear and weak shock regimes, as well as large deviations when compared to the period-based blast radius. The amplitude-based $R_0$ estimate severely under-predicts the observed amplitude in the linear regime, and overestimates it in the weak shock regime.

iii. Drawing upon the results from (ii), we varied the distortion distance to examine its effect on the weak shock to linear transition altitude. We empirically established that to match the observed amplitude of the meteor infrasound at the ground within the context of the ReVelle [1974] model, the distortion distance for our dataset must always be less than six percent, contrary to the proposed fixed 10 percent [Towne, 1967]. The choice of definition of the distortion distance has a strong effect on the predicted linear amplitude. We established the 'best fit' linear regime $R_0$ which matched both amplitude and period within their respective measurement uncertainties. No one definition of modified distortion distance worked for all events, but we found acceptable fits when the distortion distance was assumed to be of an order half or less than the original ReVelle [1974] adopted value.

iv. We applied the FM entry model [Ceplecha and ReVelle, 2005] to photometric data as measured from video observations of meteor events to independently calculate the blast radius using the fundamental definition for $R_0$ in terms of energy deposition per path length derived from the FM model. This blast radius was used as an input for top-down modelling to determine the predicted signal amplitude and period in both weak shock and linear regimes. Both the predicted period and amplitude as obtained from the best fit $R_0$ are nearly 1:1. This validates the basic definition of blast radius and its fundamental linkage to energy deposition during the hypervelocity meteor entry.

v. The infrasonic mass estimate is systematically larger than the mass estimated from FM modeling and is not a reliable predictor of the true meteoroid mass using any set of assumptions which we interpret as being mainly due to the ubiquitous presence of fragmentation during ablation of centimeter-sized meteoroids. The fragmentation tends to artificially increase the equivalent single-body mass, making infrasonically determined mass less reliable for larger events.

vi. We derived new empirical relations which link the observed dominant signal period to the meteoroid blast radius: $R_0 = 15.4 \ \tau - 0.5 \ (\tau_{obs} \leq 0.7\text{s})$ and $R_0 = 29.1 \ \tau - 11.6 \ (\tau_{obs} > 0.7\text{s})$. The blast radius can be estimated to within 15 percent.

Even though the premise of the ReVelle [1974] weak shock model is to require some knowledge about the source *a priori* to be able to predict the signal amplitude and the period at receiver located at the ground, we have obtained an empirical relation which can be used to estimate the source blast radius for centimeter-sized bright fireballs, regardless of the meteoroid's velocity, entry angle and other parameters which are generally difficult to determine without video observations. In conclusion, the weak shock model of meteor infrasound production [ReVelle, 1974] in its analytical form offers a good first order estimate in determining the blast radius and therefore energy deposition by small meteoroids, particularly if period alone is used or if no fragmentation is present.

## Acknowledgements

The authors would like to thank: J. Borovicka, P Spurny and Z. Ceplecha for kindly providing trajectory solution software and the FM model code. EAS thanks: CTBTO Young Scientist Award (through European Union Council Decision 2010/461/CFSP IV) for funding part of this project; L. Sutherland for generously providing worked out atmospheric absorption coefficients; R. Weryk for assistance with the synthetic photometry and all-sky camera development; J. Gill for setting up the parallel processing machine; the British Atmospheric Data Centre for providing UKMO Assimilated atmospheric data set. PGB thanks the Natural Sciences and Engineering Research Council of Canada, the Canada Research Program and NASA cooperative agreement NNX11AB76A for funding this work. The authors acknowledge thoughtful comments from O. Popova and one anonymous reviewer which helped improve the manuscript. The data used in this study are available upon request to the author by e-mail (contact e-mail: esilber@alumni.uwo.ca).

## Figures and Figure captions:

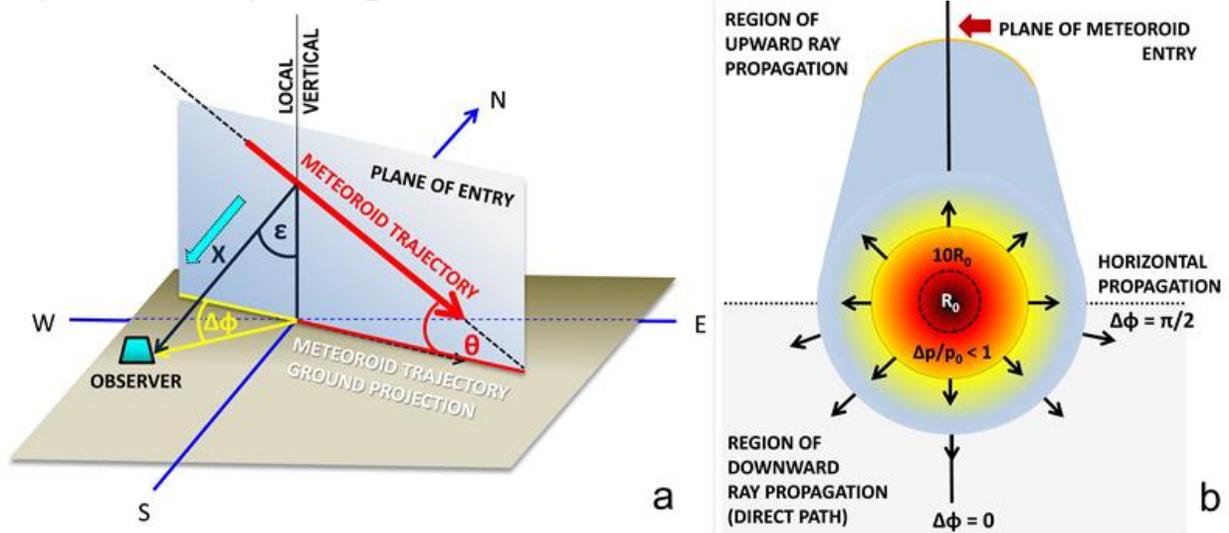

Figure 1: (a) The side view of the meteor path and (b) the cross-sectional view of the shock cavity with the meteoroid moving out of the plane of the page in the coordinate system used for modelling meteor infrasound as originally defined by ReVelle [1974]. The meteoroid trajectory is within the entry plane. The vertical plane containing the meteoroid trajectory whose normal is tangent to the Earth's surface is referred to as the plane of entry. Azimuth angles, as viewed from the top looking downward, are measured clockwise from North. The variable $x$ refers to the distance between the point along the trail and the observer in units of blast-radii. In this coordinate system, the variables are: $\varphi$ = azimuth angle of the meteoroid heading; $\varphi'$ = azimuth angle of the infrasonic wave vector outside the entry plane; $\Delta\varphi = |\varphi - \varphi'|$ = infrasound ray deviation from the plane of entry ($\Delta\varphi = 0$ ray propagation in the same azimuthal direction as the plane of entry, $\Delta\varphi = \pi/2$ out of the entry plane, i.e. purely horizontal and in the direction of the normal vector to the entry plane); $\theta$ = meteoroid trajectory entry angle as measured from the horizontal ($\theta = \pi/2$ is vertical entry); $\varepsilon$ = nadir angle of the infrasonic ray with respect to the local vertical, where $\varepsilon \geq \theta$ ($\varepsilon = 0$ is vertically downward, $\varepsilon = \theta$ in the plane of entry), always as viewed from the normal direction to the plane of entry. This angle was originally defined as zenith angle facing downward [ReVelle, 1974]; $x$ = the range ($R$) between the point along the trail and the observer in units of blast radii ($x = R/R_0$). The relationship between the zenith angle of the infrasonic ray wave vector emitted at the source which ultimately reaches the station, the meteoroid trajectory entry angle and infrasound ray deviation azimuthal angle are: $\varepsilon = \tan^{-1}\{[1 - (2\Delta\varphi/\pi)] \cot\theta\}^{-1}$ and $\Delta\phi = \pi/2 (1 - \tan\theta \cot\varepsilon)$, where $\varepsilon \neq 0$; $\theta \neq \pi/2$; $\varepsilon \geq \theta$.

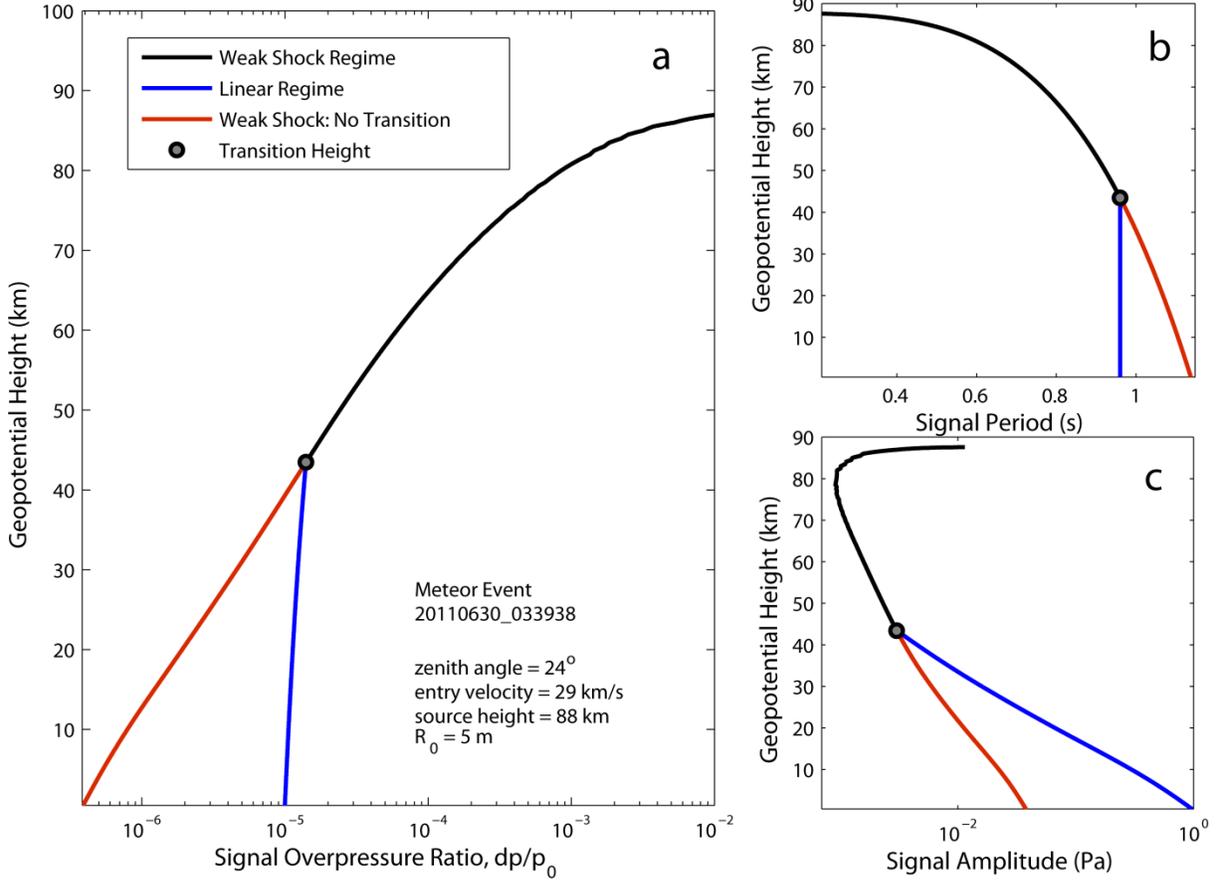

Figure 2: The change in signal (a) overpressure ratio (d*p*/*p₀*), (b) period, and (c) amplitude as a function of height from source to receiver in a fully realistic atmosphere (with winds and true temperature variations with height) according to the ReVelle [1974] theory.

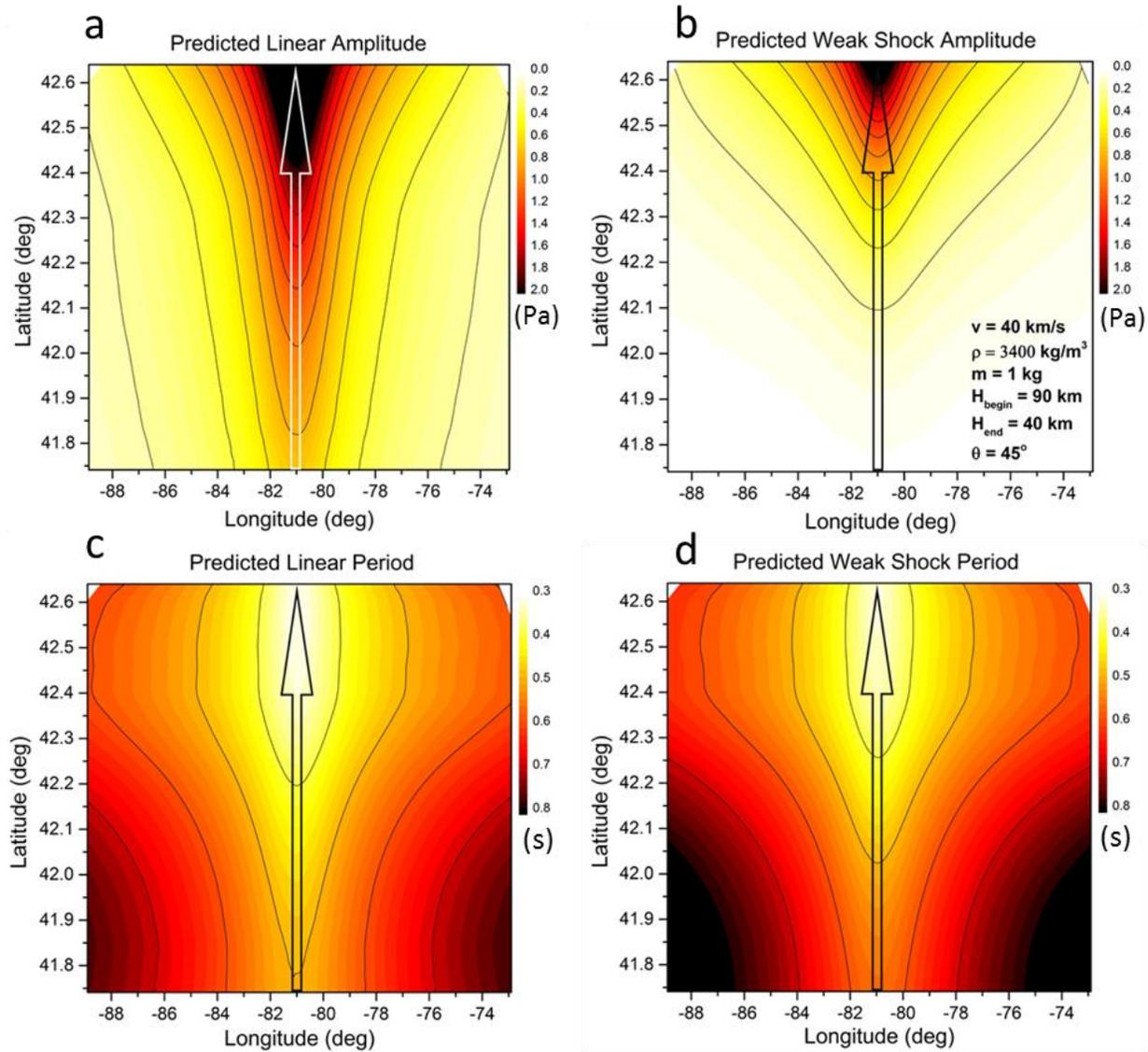

Figure 3: An example of the predicted ground-level amplitude and period of a meteor shock using the ReVelle [1974] theoretical model. In these figures, the meteor moves northward, as shown with the arrow in each plot. The meteor trajectory extends from an altitude of 90 km down to 40 km. A representative realistic atmosphere was applied, accounting for the wind. The top two panels show the predicted (a) linear and (b) weak shock amplitude. The bottom two panels show the predicted (c) linear and (d) weak shock period. The amplitude in the linear regime has a larger magnitude than that in the weak shock regime, while the opposite is true for the signal period. The synthetic meteor parameters are shown in the lower right of plot (b).

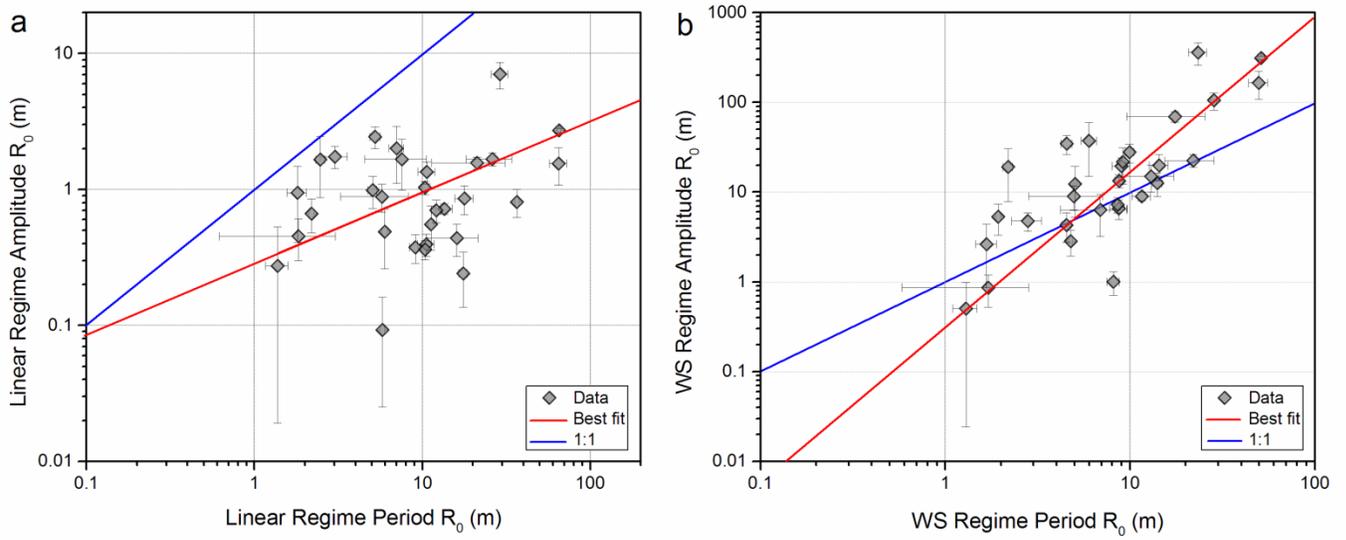

Figure 4: The behaviour of $R_0$ in (a) linear and (b) weak shock regimes in the bottom-up modelling approach.

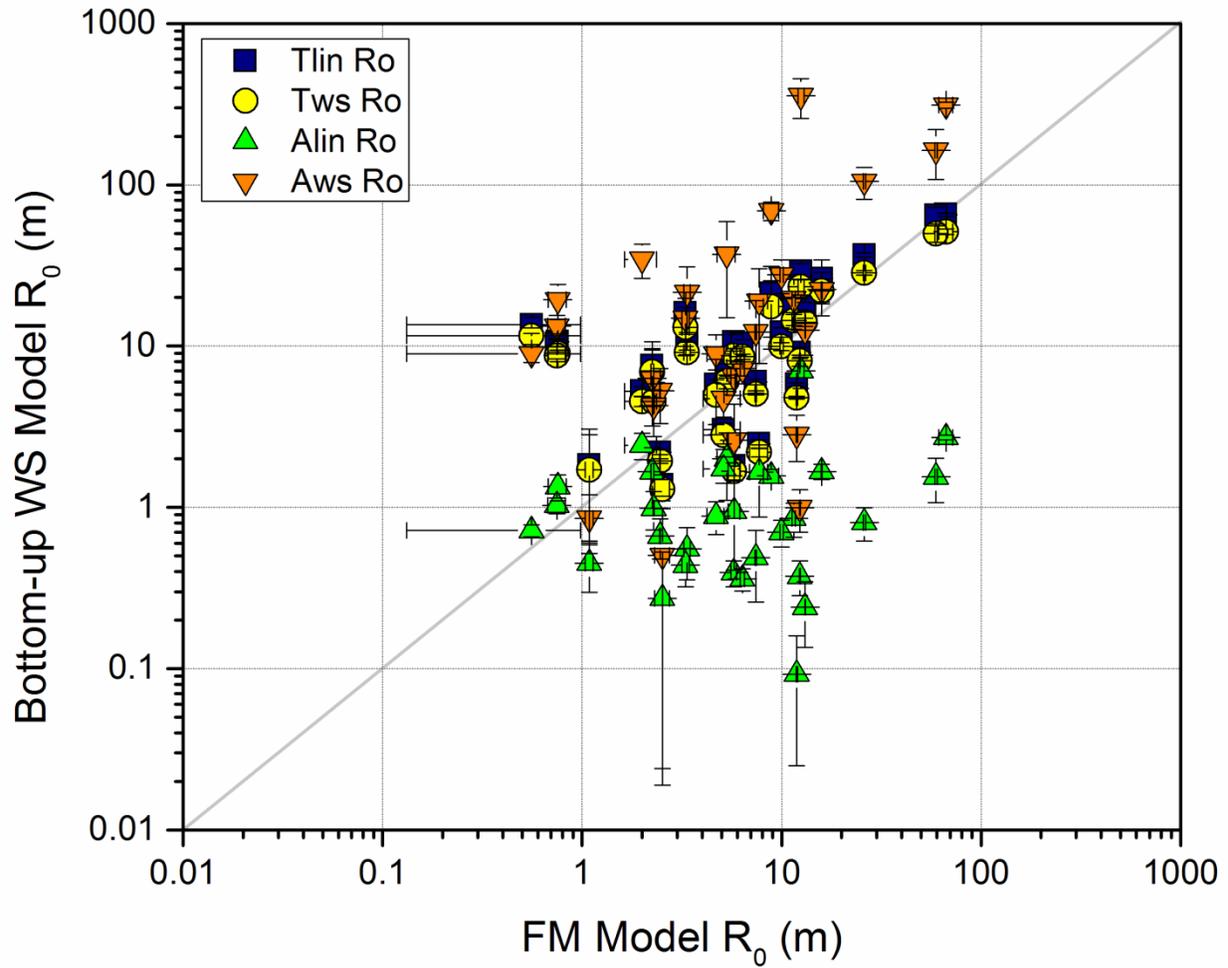

Figure 5: The blast radius estimated from bottom-up modelling as compared to blast radii derived from application of the FM model. Tlin and Tws are the period in linear and weak shock regime, respectively. Alin and Aws are the amplitude in linear and weak shock regime, respectively.

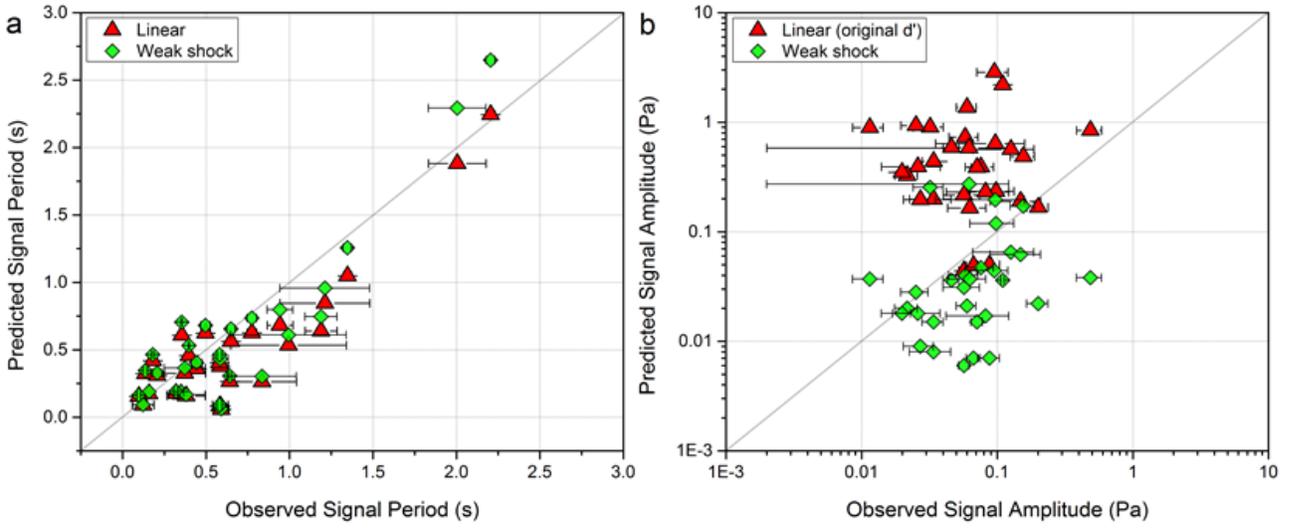

Figure 6: (a) Signal period and (b) amplitude obtained by running the top-down weak shock model with input $R_0$ as derived via the FM model.

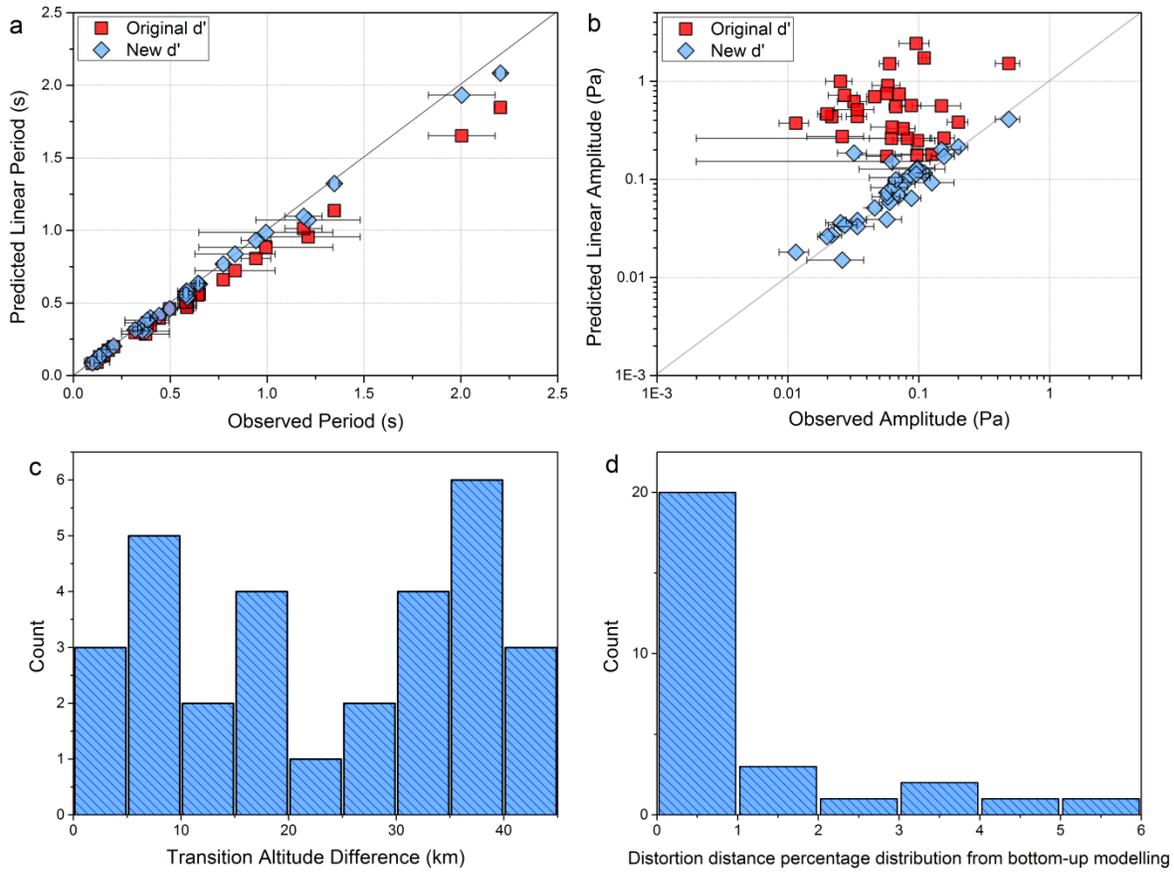

Figure 7: The best fit $R_0$ from bottom-up modelling predictions for amplitude and period vs observed (a) linear period and (b) linear amplitude. The square data points represent model output for the original definition for the distortion distance [Towne, 1967], while the diamonds represent the new values. The solid gray line is the 1:1 fit. (c) Distribution of transition altitudes for the original and new distortion distance. (d) Distribution of the transition altitude difference ($\Delta H_{transition}$) between the original and new distortion distance percentage values that were used to find the optimal $R_0$ fit; note original definition was 10%.

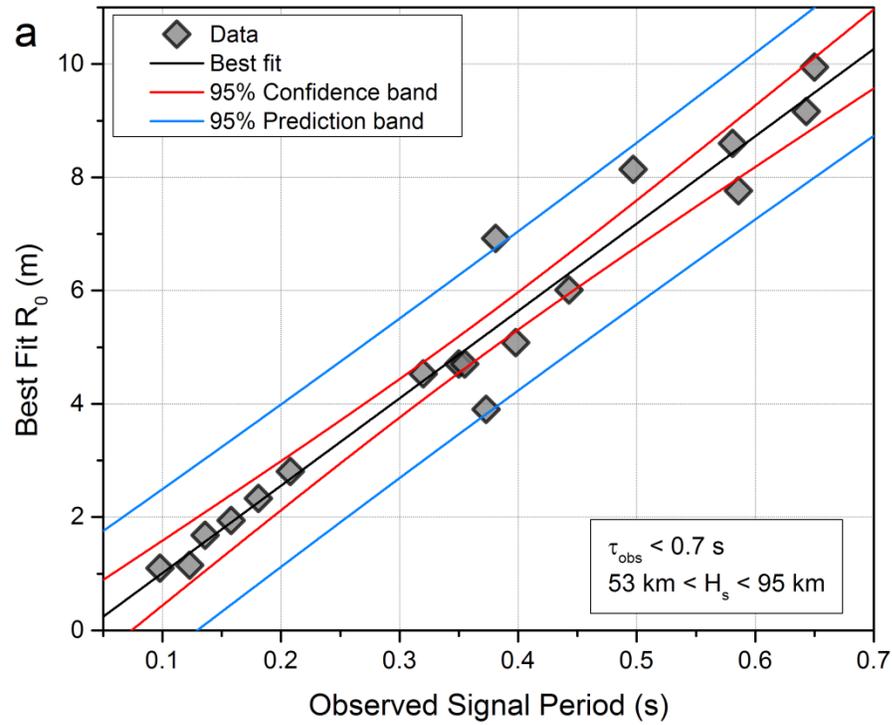

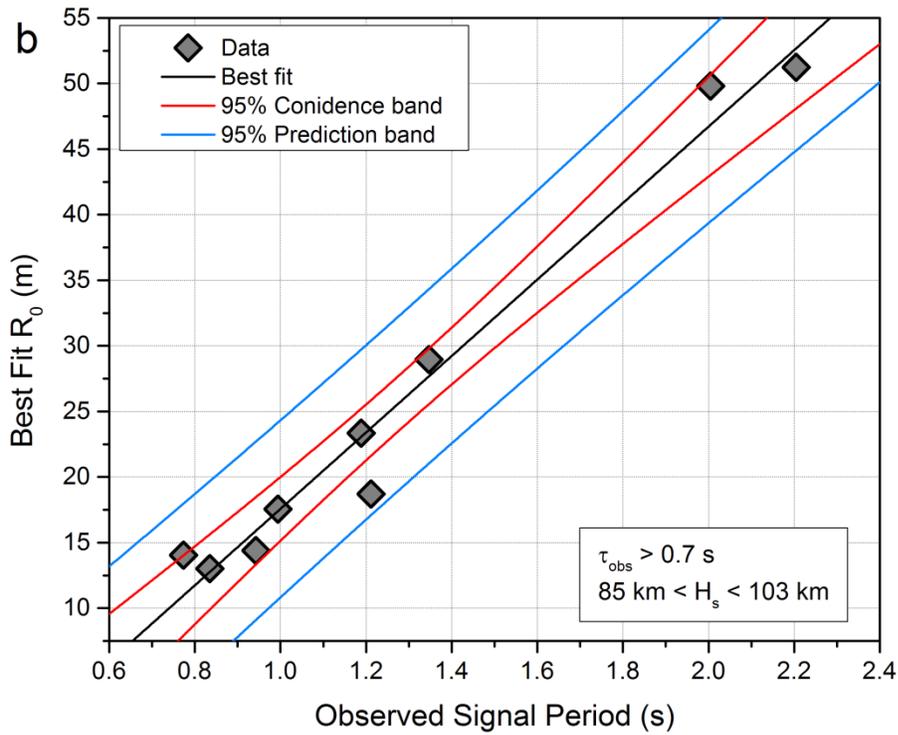

Figure 8: Blast radius from best fit bottom-up modelling versus observed signal for (a) the short signal period ($\tau_{obs} \leq 0.7$ s) and (b) the long signal period ($\tau_{obs} > 0.7$ s) populations. $H_s$ is the shock source height (Table S2) as derived from raytracing [Silber and Brown, 2014].

Auxiliary Material for

Optical Observations of Meteors Generating Infrasound - II: Weak Shock Theory and Validation

Elizabeth A. Silber

(Department of Physics and Astronomy, University of Western Ontario, London, ON N6A 3K7, Canada)

Peter G. Brown

(Department of Physics and Astronomy, University of Western Ontario, London, ON N6A 3K7, Canada)

Zbigniew Krzeminski

(Department of Physics and Astronomy, University of Western Ontario, London, ON N6A 3K7, Canada)

Introduction

This auxiliary material contains the supplemental text, figures and tables.

text01.pdf

S.1 A Modified Version of f(x)

S.2 The ReVelle Weak Shock Model Updates

S.3 Sensitivity Study

S.4 Photometric Measurements

S.5 The FM Model

Figure S1: The FM model fit to the light curve for a multi-station meteor recorded at 07:05:56 UT on 19 April 2006. The model is matched in both (a) magnitude vs. time and (b) height residuals vs. time (dynamic fit). In this case the FM model provides a best fit of 14.18 km/s initial speed and 20 g mass.

Figure S2: The comparison of the infrasonic and photometrically derived masses. The abscissa in both panels represents the equivalent photometric mass derived from the FM model blast radius and equation (8). The infrasonic masses derived from the amplitude-based (top panel) and period-based (bottom panel) blast radius in the linear and weak shock regimes are shown. The grey line is the 1:1 line in all plots.

Figure S3: Energy per unit path length comparison between the bottom-up modelling, where $R_0$ is determined entirely from the infrasound signal properties at the ground, and the FM model. The grey line is the 1:1 line.

Figure S4: The comparison of the infrasonic and photometrically derived masses. (a) Infrasonic mass from the best fit $R_0$. The three points on the far left are from the two events, which were poorly constrained in terms of the transition height and distortion distance. The abscissa represents the equivalent photometric mass derived from the FM model blast radius and equation (8); (b) The comparison between the FM model input mass and the mass derived from the blast radius as per equation (8). The grey line is the 1:1 line in all plots.

Figure S5: (a) Blast radius produced by a single spherically shaped body, with no fragmentation or strong ablation. (b) Blast radius (exaggerated depiction/schematic) produced during fragmentation and/or significant ablation.

Table S1: A summary of orbital parameters for all events in our data set. The columns are as follows: (1) event date, (2) Tisserand parameter, a measure of the orbital motion of a body with respect to Jupiter [Levison, 1996], (3) semi-major axis (AU), (4) eccentricity, (5) inclination (°), (6) argument of perihelion ($\omega$) (°), (7) longitude of ascending node (°), (8) geocentric velocity (km/s), (9) heliocentric velocity (km/s), (10) $\alpha$ – right ascension of geocentric radiant (°), (11) $\delta$ – declination of geocentric radiant (°), (12) perihelion (AU), (13) aphelion (AU), (14) meteor shower associations. The meteor shower codes are: $\alpha$-Capricornids (CAP), Orionids (ORI), Perseids (PER), and Southern Taurids (STA). Note all angular quantities are J2000.0.

Table S2: A summary of the event parameters. The table columns are as follows: (1 – 4) event date and time, (5 – 6) the shock source height and its uncertainty, (7) total range[*] and (8) horizontal range[*], (9) meteor azimuth, (10) meteor trajectory length, (11) meteor flight time, (12) observed infrasonic signal travel time from the point of shock to the receiver, (13 – 14) observed infrasound signal back azimuth with its uncertainty, (15) measured infrasound signal dominant frequency, (16 – 17) measured infrasound signal dominant period with its uncertainty, (18 – 19) maximum signal amplitude with its uncertainty and (20 – 21) peak-to-peak amplitude with its uncertainty. The details outlining the methodology as to how these parameters were determined can be found in Silber and Brown [2014]. [*]Range denotes the distance in km from the point of shock along the meteor trajectory to the infrasound station.

Table S3: A summary of the photometric mass measurements using observed light curves from meteor video records. The columns are as follows (column numbers are in the first row): (1)

meteor event date, (2) meteor velocity (km/s) at the onset of ablation, (3) begin height (km), (4) end height (km), (5) instrument corrected peak magnitude ($M_{HAD}$) from the observed light curve, (6-7) minimum and maximum values of shape coefficient ($K$), and (8-9) minimum and maximum values of ablation coefficient ($\sigma$) used in the FM model fitting, (10) estimated meteor mass (g) from equation (8), (11) estimated meteor mass (g) using integrated light curve, (12) initial (entry) meteor mass (g) from the FM model, (13) PE coefficient and (14) meteoroid group type.

Table S4: The range of PE parameter and the associated meteoroid group presumed type, type of the material and representative density as given by Ceplecha et al. [1998]. The percentage of observed meteoroids in each group according to mass category typical of fireball networks (mass range 0.1 - 2000 kg) and small cameras (mass range $10^{-4}$ kg to 0.5 kg) as published in [Ceplecha et al., 1998] and the percentages found in this study (masses $10^{-4}$ kg to 1.0 kg). The range of values for the intrinsic ablation coefficients ($\sigma$) and shape density coefficient ($K$) found in our study within the framework of the FM model are given in the last two columns.

Table S5: A summary of $R_0$ as derived from the FM model. The associated outputs from top-down modelling are also shown. The two entries in the Taltitude, which is the transition height, in the "New d'" section denote events that never transition to the linear regime. Therefore, the original definition for d' was used by default. [*]There was no convergent solution.

**Auxiliary Material: Silber et al. (2015), Optical Observations of Meteors Generating Infrasound - II: Weak Shock Theory and Validation**

## Table of Contents

## S.1 A modified Version of f(x)

The function *f(x)* (equation (4a)) can be slightly modified using constants $Y_C$ and $Y_D$ based on the work of Plooster [1968]:

$$f(x) = \left(\frac{3}{8}\right)^{-3/5} Y_C^{-8/5} \left\{ \left[ 1 + \left(\frac{8}{3}\right)^{8/5} Y_C^{-8/5} Y_D^{-1} x^2 \right]^{3/8} - 1 \right\}^{-1} \quad (S1)$$

$Y_C$ is Plooster's adjustable parameter [ReVelle, 1976] which defines the region where the nonlinear to weak shock transition occurs, while $Y_D$ describes the efficiency with which cylindrical blast waves are generated as compared to the results of an asymptotic strong shock as numerically determined by Lin [1954]. A high value of $Y_D$ implies that the rate of internal energy dissipation is low, thereby leaving more energy available for driving the leading shock [Plooster, 1968]. The choice of $Y_C = Y_D = 1$ as implemented by ReVelle [1974] leads to equation (4b), where $\Delta p = 0.0575 p(z)$ at $x = 10$. While the value of $\Delta p$ (at $x = 10$) may be as high as $0.0906 p(z)$ depending on the choice of $Y_C$ and $Y_D$ as experimentally determined by Plooster [1968], by the time the wave reaches the receiver, the difference in the final overpressure found using the extreme value of the terms proposed by Plooster [1968] is less than 1% [Silber, 2014]. As a result, we do not expect to be able to constrain these terms using meteor infrasound observations at the ground, nor are the magnitude of the differences significant in comparison to other sources of uncertainty and hence we adopt the same values used by ReVelle [1974].

## S.2 The ReVelle [1974] Weak Shock Model

For shock amplitudes, Morse and Ingard [1968] showed that the effects of nonlinear terms compared to viscous terms are not negligible until $\Delta p$ is sufficiently small that the atmospheric gas mean free path becomes much greater than displacements of particles due to wave motion. Thus, the following equation applies to 'shocked' acoustic waves [Morse and Ingard, 1968] at distances far from the source:

$$\frac{dp_s}{ds} = -\left(\frac{\gamma + 1}{\gamma^2 \lambda}\right)\left(\frac{\rho_0 c^2}{p_0^2}\right) p_s^2 - \left(\frac{3\delta}{2\rho_0 c\lambda^2}\right) p_s \quad (S2)$$

where $p_s$ is the pressure amplitude of the 'shocked' disturbance, $ds$ is the path length, $\lambda$ is the wavelength of the 'shocked' disturbance, $\rho_0$ is the ambient air mass density (assumed to be approximately the same as the shocked air density), $p_0$ is the ambient air pressure, and $\delta$ is defined as:

$$\delta = 4\left[\frac{4}{3}\mu + \psi + K_0\left(\frac{\gamma - 1}{C_P}\right)\right] \qquad (S3)$$

Here $\mu$ is the ordinary (shear) viscosity coefficient of the medium, $\psi$ is the bulk (volume) viscosity coefficient, $K_0$ is thermal conductivity of the fluid, and $C_P$ is the specific heat of the fluid at constant pressure. The term $p_s{}^2$ in equation (S2) is due to viscous and heat conduction losses across entropy jumps at the shock fronts, while $p_s$ is due to the same mechanisms, but between the shock fronts [ReVelle, 1974]. A more compact version of equation (S2), represented in terms of coefficients A and B is given by:

$$dp_s = -(Ap_s^2 + Bp_s)ds \quad (S3)$$

where

$$A(z) = \frac{\gamma + 1}{\gamma\lambda(z)p_0} \quad (S4a)$$

and

$$B(z) = \frac{3\delta}{2\rho_0(z)c(z)\lambda(z)^2} \quad (S4b)$$

It is assumed that $\gamma H \gg \lambda(z)$ and $\rho_0(z)c(z)^2 = \gamma p_0$. Here, $H$ is the scale height of the atmosphere and $\lambda$ is the wavelength of the shock. The latter approximation is valid if the density ratio across the shock front is not significantly above unity, which is applicable in the regime far from the source, where $\rho/\rho_0 \leq 1.5$ (for $x \geq 1$, Plooster [1968]) where our measurements are made.

The generalized form of the damping factor, or atmospheric attenuation, for a weak shock [ReVelle, 1974] is given by:

$$D_{ws}(z) = \frac{\Delta p_t}{\Delta p_z} = \frac{\frac{B(z)}{A(z)}exp\left(-\int_{z_z}^{z_t}\frac{B(z)}{cos\,\varepsilon}dz\right)}{\Delta p_z\left[1 - exp\left(-\int_{z_z}^{z_t}\frac{B(z)}{cos\,\varepsilon}dz\right)\right] + \frac{B(z)}{A(z)}} \qquad (S5)$$

where $\Delta p_z$ is the overpressure at the source height, $\Delta p_t$ is the overpressure at the transition height $z_t$ where $d' > d_a$ is satisfied. Beyond the transition height where propagation is assumed to be linear, the absorption decay law for a plane sinusoidal wave, as given by Evans and Sutherland [1971] takes the form:

$$\frac{\Delta p}{\Delta p_z} = \exp(-\overline{\alpha}\Delta s) \qquad (S6)$$

where $\Delta s$ is the total path length from the source ($\Delta s \geq 0$), and $\overline{\alpha}$ is the total absorption amplitude coefficient [Morse and Ingard, 1968], which can be written as:

$$\overline{\alpha}(z) = \frac{\omega^2}{2\rho(z)c(z)^3}\left[\frac{4}{3}\mu + \psi + K\left(\frac{\gamma - 1}{C_P}\right)\right] = \frac{\pi^2}{\rho(z)c(z)\lambda(z)^2}\left(\frac{\delta}{2}\right) \qquad (S7)$$

Here, $\omega$ is the angular frequency of the oscillation ($\omega = 2\pi f$).

The generalized form of ReVelle's [1976] damping function for a linear wave is given by:

$$D_l(z) = \frac{\Delta p_{obs}}{\Delta p_t} = exp\left(-\int_{z_t}^{z_{obs}}\frac{\overline{\alpha(z)}}{cos\ \varepsilon}dz\right) \qquad (S8)$$

Note that the integration limits in equation (S8) are from the transition height (where the shock is assumed to change from weak shock-to-linear propagation) down to the receiver.

In addition to the damping coefficients ($D_{ws}$ and $D_l$), additional correction terms are required to account for linear propagation, non-uniform refracting path and density variations along the propagation path:

1. In the linear propagation regime the asymptotic form of the wave overpressure decay goes as $x^{-1/2}$ [Officer, 1958] compared to $x^{-3/4}$ in the weak shock regime.

2. A correction term for amplitude to account for differences between the actual refractive path to that of a straight-line path between the source and the receiver [Pierce and Thomas, 1969]:

$$N^*(z) = \left(\frac{\rho(z)}{\rho_z}\right)^{\frac{1}{2}}\frac{c(z)}{\overline{c_z}}N_c \qquad (S9)$$

$$\overline{c_z} = \frac{\int_{z_{obs}}^{z_z} c(z)dz}{z_z - z_{obs}} \quad (S10)$$

where $N_c$ is the nonlinear propagation correction term, assumed to be unity [Cotten et al., 1971; ReVelle, 1974], $c(z)$, $\rho(z)$ are the sound speed and atmospheric density as a function of altitude between the source and the receiver, $z_{obs}$ is the altitude (0 if at the ground) at the receiver, respectively, and $\rho_z$ and $z_z$ are the atmospheric density and altitude at the source, while $\overline{c_z}$ is the average speed of sound between the source and the receiver. In general this term takes on values below 2 and can be neglected compared to other uncertainties associated with such processes as turbulence and local reflections near the ground not explicitly taken into account in the amplitude determination.

3. An atmospheric density correction term [Pierce and Thomas, 1969; ReVelle, 1974] is applied to equation (4a):

$$Z^*(z) = \frac{\rho_z}{\rho(z)}\left(\frac{c_z}{c(z)}\right)^2 \quad (S11)$$

This atmospheric density correction term is the ratio of acoustic impendence at the source altitude to the receiver altitude and serves to correct the amplitude for the differences between the two heights. In deriving the overpressure ratio decay *(Δp/p)*, Jones et al. [1968] assumed a uniform ambient pressure but since meteoroids occur at high altitudes, the source pressure is often a small percentage of the receiver pressure and this term becomes significant.

ReVelle's [1974] model as just described was also recently updated to include a realistic atmosphere [Edwards et al., 2008], albeit without inclusion of the winds. In this study, however, we expanded the model further to include the effect of atmospheric winds on the signal period (Doppler shift) and amplitude; and updated absorption coefficients (modification to equation (S7)) [Sutherland and Bass, 2004], with an option to turn these on or off.

## S.3 The ReVelle Weak Shock Model Updates

The ReVelle [1974] weak shock model algorithm was implemented in *MATLAB*® and updated to include full wind dependency, as well as Doppler shift for period [Morse and Ingard, 1968] as a function of altitude.

The influence of the winds is reflected in the effective speed of sound ($c_{eff}$), which is given by the sum of the adiabatic sound speed ($c$) and the dot product between the ray normal ($\boldsymbol{n}$) and the wind vector ($\boldsymbol{u}$): c$eff = c + \boldsymbol{n}\cdot\boldsymbol{u}$.

The signal amplitude is affected by winds such that the amplitude will intensify for downwind propagation and diminish in upwind propagation [Mutschlecner and Whittaker, 2010]. In the linear regime, the signal period ($\tau(x) = 0.562\ \tau_0\ x^{1/4}$), does not suffer any decay with distance, but the winds do induce a Doppler shift. Following Morse and Ingard [1968], the Doppler shift due to the wind is given by: $\Omega = \omega - \boldsymbol{k}\cdot\boldsymbol{u}$, where $\Omega$ is the intrinsic angular frequency (frequency in the reference frame of the moving wind with respect to the ground), $\omega$ is the angular frequency in the fixed earth frame of reference and $\boldsymbol{k}$ is the wave number. Since the contribution of winds in the vertical direction is generally 2-4 orders of magnitude smaller than the horizontal wind contribution [Wallace and Hobbs, 2006; Andrews, 2010], it is neglected.

Another addition to the weak shock model was the inclusion of updated absorption coefficients [Sutherland and Bass, 2004], applicable in the linear propagation regime.

The calculation of the predicted shock signal properties (signal overpressure and period) at each altitude increment (divisions of $0.01 - 0.02$ km) were done such that the winds or newer attenuation coefficients could be turned on or off, a feature proven very useful for model testing.

The outputs of the weak shock model were tested using a synthetically generated meteor of roughly chondritic density ($\rho = 3400$ kg/m$^3$), with mass of 1 kg, entry angle of 45°, and velocity of 40 km/s, beginning ablation at the altitude of 90 km and reaching a terminal point at the altitude of 40 km. Since the weak shock model takes only a single height as an input, the entire trajectory was divided into a number of discrete heights, each serving as a separate shock height input for the model. The model was then run with a realistic atmosphere for each input height, while keeping all other input parameters constant and the resultant signal overpressure and period in both weak shock and linear regime recorded. The amplitude and period fields for this validation run are described in the results section in the main text.

## S.4 Sensitivity Study

In the first part of our study [Silber and Brown, 2014] we described the influence on the raytracing results of small scale perturbations in the wind profile due to gravity waves on raytracing results. While the effects were small, they were significant enough to produce propagation paths which were non-existent using the average atmosphere from the source to the receiver. These perturbations also led to larger uncertainty bounds in the source shock height. Here we have used the same 'perturbed' atmospheric profiles to test the influence of gravity-wave-induced perturbations on the predicted signal amplitude and period as calculated using the weak shock model. Having estimated the model blast radii in the previous step, we selected five events which span the global range of our final data (i.e. meteors with different entry velocity, blast radius, and shock heights), and ran the weak shock code using 500 'perturbed' atmospheric profiles. For each event and each realization we computed the magnitude of the modelled infrasonic signal period and amplitude while simultaneously testing the effect of different absorption coefficients in the linear regime using the set given by ReVelle [1974] and that of Sutherland and Bass [2004].

The linear amplitude-derived $R_0$ was on average 12% larger when the Sutherland and Bass [2004] coefficients were applied. To put this in perspective, the range of $R_0$ from matching the amplitude in the linear regime was from 0.15 m to 7.4 m with the Sutherland and Bass [2004] coefficients and from 0.09 m to 7 m using the classical coefficients [ReVelle, 1974]. The most significant difference was found for signals with a dominant frequency > 12 Hz (at the weak shock to linear regime transition altitude point). However, even in this case, the difference is still within the uncertainty bounds in $R_0$ fits.

## S.5 Photometric Measurements

The fundamental equations of motion [Bronshten, 1983; Ceplecha et al., 1998] for a meteoroid of mass $m$ and density $\rho_m$ entering the Earth's atmosphere at velocity $v$ are given by the drag and mass-loss equations:

$$\frac{dv}{dt} = -\frac{\Gamma A \rho_a v^2}{m^{\frac{1}{3}} \rho_m^{\frac{2}{3}}} \qquad (S12)$$

$$\frac{dm}{dt} = -\frac{\Lambda A \rho_a v^3 m^{\frac{2}{3}}}{2\xi \rho_m^{\frac{2}{3}}} \qquad (S13)$$

Here, $\Gamma$ is a drag coefficient, $\xi$ is the heat of ablation of the meteoroid material (or energy required to ablate a unit mass of the meteoroid), $\Lambda$ is the heat transfer coefficient, which is a measure of efficiency of the collision process in converting kinetic energy into heat [McKinley, 1961], $\rho_a$ is density of the air, and $A$ is the dimensionless shape factor ($A_{sphere} = 1.209$).

The all-sky meteor camera system used for this survey and details of the astrometric reductions and measurements methodology are presented in Silber and Brown [2014]. Here we briefly describe the photometric analysis.

A series of laboratory experiments were performed to determine the camera response in an effort to model the effects of camera saturation (which affect many of our events). In the field, measurements were performed to model the lens roll-off (the apparent drop in sensitivity of the camera as the edges of the all-sky field are reached). The lens roll-off correction becomes significant for meteors as a >0.8 astronomical magnitude attenuation results for elevations below ~15 degrees. Other standard photometric corrections were applied including extinction correction [e.g. Vargas et al., 2001; Burke et al., 2010] and apparent instrumental magnitude converted to an absolute magnitude by referencing all meteors to a standard range of 100 km [McKinley, 1961].

Bright meteors (M > -4) are typically saturated on the 8-bit cameras. In our laboratory setup, an artificial star of variable brightness was created using a fixed light source and a turning wheel with a neutral density filter of varying density following the procedure used by Swift et al. [2004]. Standard photometric procedures [Hawkes, 2002] were used to determine the apparent instrumental magnitude of the artificial star and a power law fit between the observed and known brightness of the artificial star was then computed to find a correction for the saturation which we applied to meteors to deduce their true apparent magnitude.

The instrumental magnitude of any given star (or meteor) varies as a function of distance from the optical center (or zenith distance if the camera is vertically directed) of the camera lens. For our cameras a star appears about 2.5 stellar magnitudes dimmer near the horizon than at the zenith due to the natural vignetting in the optical system. The in-field experiment was performed on a clear night by setting up the camera on a turntable attached to a fixed frame and taking a series of video frames starting from the horizon and sweeping through an angle of 180° through the zenith. Several bright stars then had their instrumental magnitudes computed as a function of distance from the optical axis to compute the lens roll-off which was found to functionally behave as $\cos^4\theta$. Applying all these corrections we then followed standard photometric routines [Hawkes, 2002], to compute the light curves for each meteor in our instrumental passband. Our cameras use hole accumulation diode CCD chips [Weryk and Brown, 2013]. The CCDs have both a wide passband and high QE making them extremely sensitive in low-light conditions. Our limiting meteor sensitivity is approximately $M_{HAD}$ = -2 corresponding to meteoroids of ~5 g or roughly 1cm in diameter at 30 km/s.

There are two approaches which can be used to determine the mass of a meteoroid given optical data. First, we may appeal to empirical estimates of the magnitude-mass-speed determined from earlier photographic surveys (e.g. [Jacchia et al., 1967]) which yields:

$$M_{HAD} = 55.34 - 8.75 \log v - 2.25 \log m - 1.5 \log \cos Z_R \qquad (S14)$$

where $M_{HAD}$ is the absolute magnitude in the HAD bandpass, m is the mass of the meteoroid in grams, $Z_R$ is the zenith angle of the radiant and where the meteoroid velocity ($v$) is expressed in cgs units. The original relation was computed in the photographic bandpass. We estimated the color correction term between our instrumental system and the photographic system, ($M_{HAD} = M_{Ph}$+1.2) by assuming each meteor radiates as a 4500K blackbody and computing the energy falling into our HAD bandpass as compared to the photographic bandpass following the synthetic photometry procedure described in Weryk and Brown [2013].

A second approach to estimate photometric mass is to directly integrate the light emission (I) of the meteoroid throughout the time of its visibility:

$$m = \frac{2 \int I dt}{\tau_I v^2} \qquad (S15)$$

where $\tau_I$ is the luminous efficiency defined as the fraction of the meteoroid's kinetic energy produced converted into radiation [Ceplecha et al., 1998]. More about this approach can be found in Ceplecha et al. [1998] and Weryk and Brown [2013].

The magnitude of the blast radius at any point along the meteor trail can be thought of as a 'snapshot' of the energy per unit path length deposited into the atmosphere by a meteoroid. This may also be equated to the meteoroid mass times the Mach number at that point assuming single body ablation [ReVelle, 1976].

While the single body approach is much more simple if the source region along the trail where the shock observed at the microphone location on the Earth's surface is produced close to the end of the meteor trail the total initial photometric mass poorly represents the actual energy/mass at the source location. If the initial mass is used to test the weak shock model, it produces erroneous results as the actual mass (and energy deposited per unit path length) would be much smaller as the initial meteoroid mass is much larger than the remnant mass near the end of the trail.

A third approach to estimating mass using optical data is to apply an entry model code to fit the observed brightness and length vs. time measured for the meteor. Here we use the fragmentation model (FM) of meteoroid motion, mass loss and radiation in the atmosphere [Ceplecha and ReVelle, 2005]. The FM model includes model estimates for the luminosity, ablation, and fragmentation which can be fit to the observed brightness and length vs time to produce estimates for mass as a function of height. The FM model should provide a more realistic estimate of energy deposition along the trail [Ceplecha and ReVelle, 2005], as it explicitly accounts for fragmentation.

## S.6 FM Model

The first step in constructing an entry model solution is to begin with the approximate (starting) values of intrinsic shape density coefficient ($K$) and intrinsic ablation coefficient ($\sigma$), which we then modify together with mass in a forward modeling process. These two quantities are defined by the following expressions [Ceplecha and ReVelle, 2005]: $\sigma = \Lambda/(2\Gamma\xi)$ and $K = \Gamma A/\rho^{2/3}$, where $\Gamma$ is a drag coefficient, $\xi$ is the heat of ablation of the meteoroid material, $\Lambda$ is the heat transfer coefficient, $\rho$ is meteoroid density, and $A$ is the dimensionless shape factor ($A_{sphere} = 1.209$). We do this statistically by first classifying each of our meteors according to the fireball types defined by Ceplecha and McCrosky [1976] (hereafter CM). This is done by calculating the PE parameter [Ceplecha and McCrosky, 1976], defined as:

$$\text{PE} = \log \rho_E + \text{A}_0 \log m + \text{B}_0 \log v + \text{C}_0 \log \cos Z_R \qquad \text{(S16)}$$

where $\rho_E$ is the density of air (in cgs units) at the trajectory end height, m is the estimated initial mass in grams, and the meteoroid entry velocity ($v$) is expressed in km/s. From fits to a large suite of photograhphically observed fireballs, CM found $A_0$ = -0.42, $B_0$ = 1.49 and $C_0$ = -1.29. As an estimate for the initial mass we use the mass computed from equation (S15) and assume a +1.2 color term between the HAD and photographic bandpasses. In this way we derive estimates for the PE for each of our events. The range of values of PE and presumed corresponding meteoroid types as proposed by CM are given in Table S3.

As our average meteoroid mass is intermediate between fireball and small camera data, we *a priori* expect our distribution of fireball types (I-III) to be intermediate between these two classes, if our mass scale is reasonable. As shown in Table S3, within the limitations of our small number statistics, our distribution is broadly consistent with being intermediate between the percentage distribution of these two categories given by Ceplecha et al. [1998] indicating our choice of color term is physically reasonable.

Using the results for 15 bolides obtained by Ceplecha and ReVelle [2005] as a starting reference point, the following parameters were forward modeled until a best fit to the observed light curve as well as magnitude as a function of height and path length were found: $K$, $\sigma$, integration altitude, the initial mass, the mass loss at each fragmentation point (if applicable), duration of the flare, interval from the start of the flare until the flare maximum and part of mass fragmented as large fragments. The ranges of input values for $\sigma$ and $K$ as found by Ceplecha and RVelle [2005] are given in Table S3, while the ranges used in this study are in Table S2.

The FM model fitting was performed using the dynamic and light curve meteor data from optical measurements. The average observed light curve to be fitted is comprised of light curves from each individual site for a given meteor event. Overall uncertainties in the light curves for our data set are relatively small (on average <0.5 magnitude), though larger uncertainties occur. These include meteors which occurred during nights with poor sky conditions (e.g. haze, thin clouds, light pollution, full Moon), which in turn may affect the photometric solution.

The input parameters for the FM model are the meteor starting height ($h_b$), geographic latitude at that point, meteoroid entry velocity ($v$), meteoroid entry angle ($Z_R$), initial mass, intrinsic shape density coefficient ($K$) and intrinsic ablation coefficient ($\sigma$). From the FM model we have the following to match to the meteor: time, path length, altitude, velocity, *dv/dt*, mass, *dm/dt*, meteor luminosity, meteor magnitude, $\sigma$, $K$, luminous efficiency and zenith distance of the radiant.

The resultant model light curve fit (magnitude vs time) includes contributions from the main body, as well as mass lost during fragmentation (i.e. flares).

**Supplemental Figures**

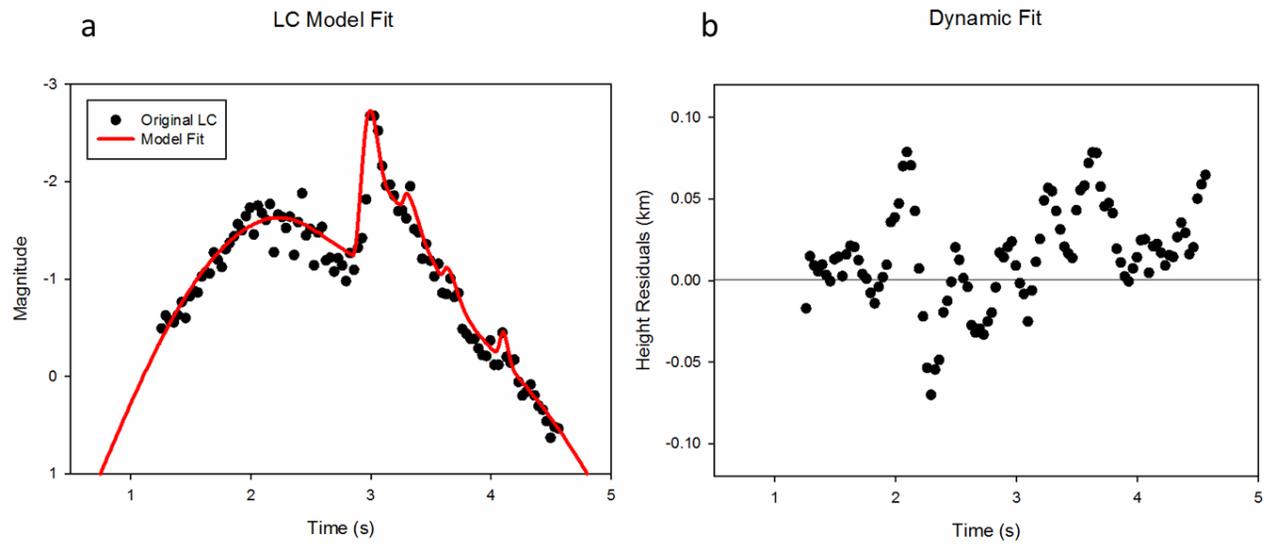

Figure S1: The FM model fit to the light curve for a multi-station meteor recorded at 07:05:56 UT on 19 April 2006. The model is matched in both (a) magnitude vs. time and (b) height residuals vs. time (dynamic fit). In this case the FM model provides a best fit of 14.18 km/s initial speed and 20 g mass.

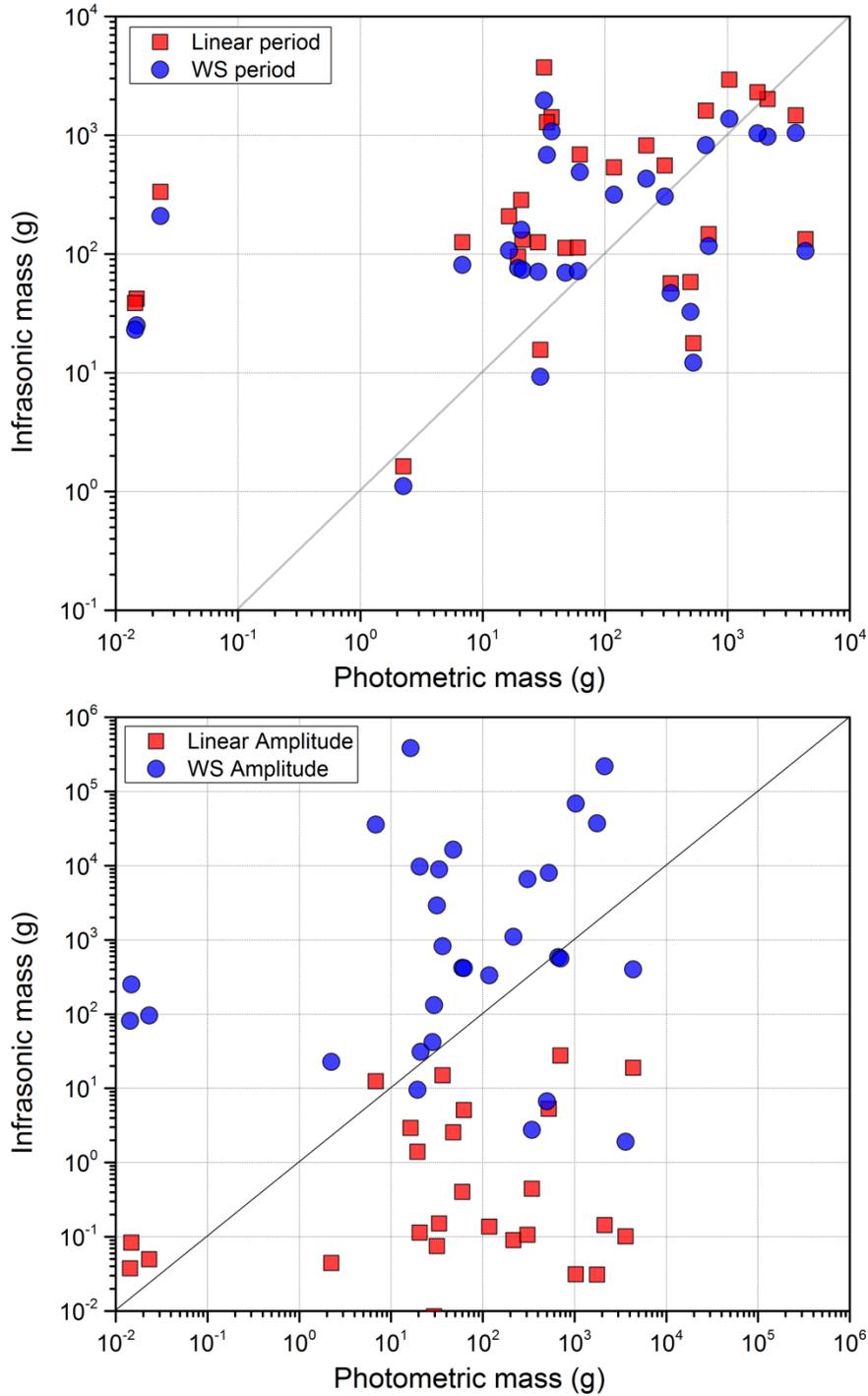

Figure S2: The comparison of the infrasonic and photometrically derived masses. The abscissa in both panels represents the equivalent photometric mass derived from the FM model blast radius and equation (8). The infrasonic masses derived from the amplitude-based (top panel) and period-based (bottom panel) blast radius in the linear and weak shock regimes are shown. The grey line is the 1:1 line in all plots.

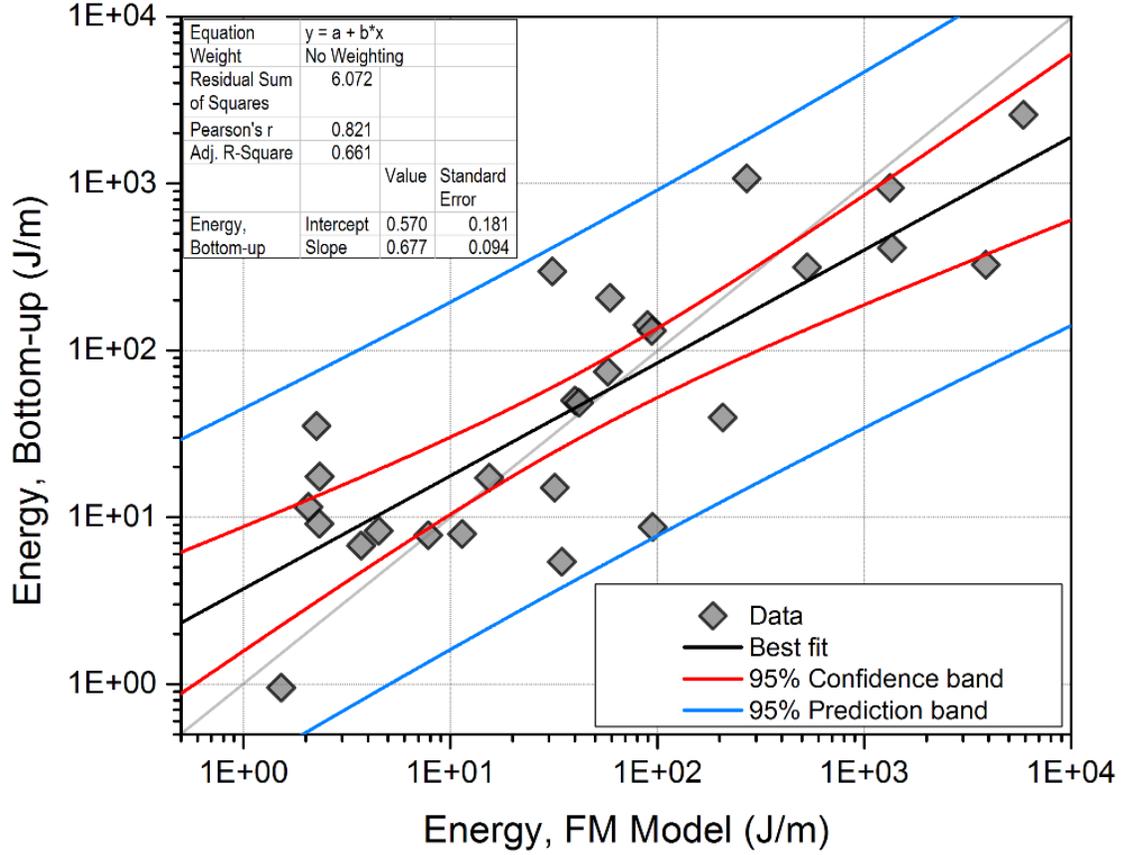

Figure S3: Energy per unit path length comparison between the bottom-up modelling, where $R_0$ is determined entirely from the infrasound signal properties at the ground, and the FM model. The grey line is the 1:1 line.

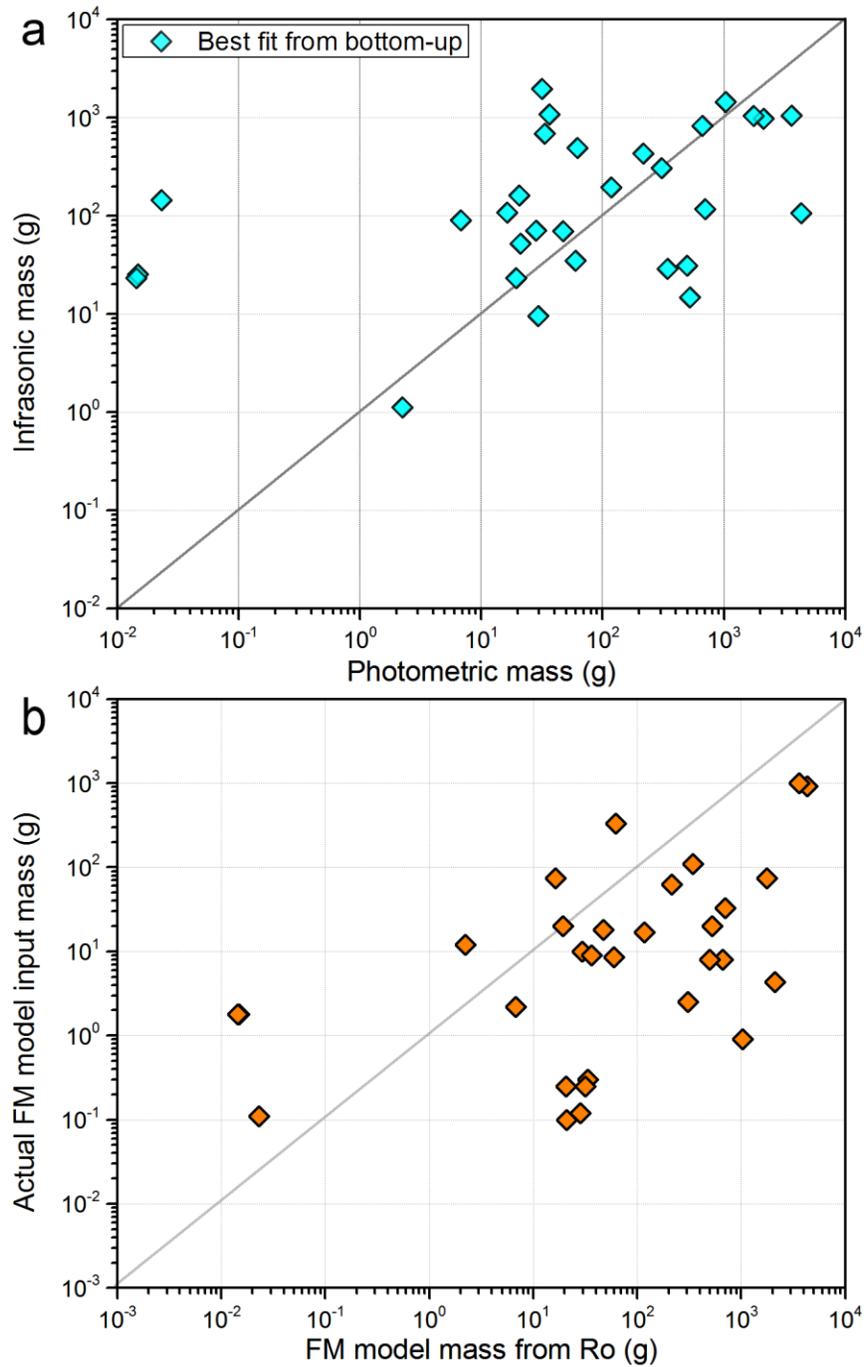

Figure S4: The comparison of the infrasonic and photometrically derived masses. (a) Infrasonic mass from the best fit $R_0$. The three points on the far left are from the two events, which were poorly constrained in terms of the transition height and distortion distance. The abscissa represents the equivalent photometric mass derived from the FM model blast radius and equation (8); (b) The comparison between the FM model input mass and the mass derived from the blast radius as per equation (8). The grey line is the 1:1 line in all plots.

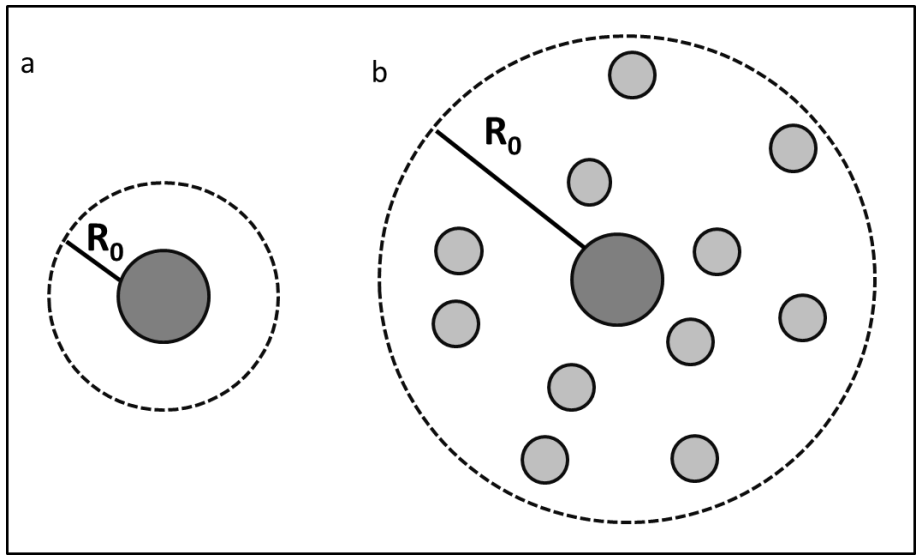

Figure S5: (a) Blast radius produced by a single spherically shaped body, with no fragmentation or strong ablation. (b) Blast radius (exaggerated depiction/schematic) produced during fragmentation and/or significant ablation.

## Supplemental Tables

Table S1: A summary of orbital parameters for all events in our data set. The columns are as follows: (1) event date, (2) Tisserand parameter, a measure of the orbital motion of a body with respect to Jupiter [Levison, 1996], (3) semi-major axis (AU), (4) eccentricity, (5) inclination (°), (6) argument of perihelion ($\omega$) (°), (7) longitude of ascending node (°), (8) geocentric velocity (km/s), (9) heliocentric velocity (km/s), (10) $\alpha$ – right ascension of geocentric radiant (°), (11) $\delta$ – declination of geocentric radiant (°), (12) perihelion (AU), (13) aphelion (AU), (14) meteor shower associations. The meteor shower codes are: α-Capricornids (CAP), Orionids (ORI), Perseids (PER), and Southern Taurids (STA). Note all angular quantities are J2000.0.

| 1 Date | 2 Tiss | 3 a | 4 e | 5 inc | 6 $\omega$ | 7 asc node | 8 $v_g$ | 9 $v_h$ | 10 α geo | 11 δ geo | 12 q per | 13 q aph | 14 Showers |
|---|---|---|---|---|---|---|---|---|---|---|---|---|---|
| 20060419 | 2.8 | 3.2 | 0.69 | 2.2 | 195.5 | 29.021 | 10.0 | 38.6 | 153.0 | 20.2 | 0.989 | 5.4 | -- |
| 20060805 | hyp | -8.4 | 1.12 | 144.9 | 164.9 | 132.716 | 69.0 | 43.1 | 38.7 | 37.1 | 0.996 | -17.7 | -- |
| 20061104 | 3.1 | 2.2 | 0.83 | 4.3 | 292.1 | 221.472 | 28.0 | 37.3 | 49.0 | 22.0 | 0.371 | 4.1 | -- |
| 20070125 | hyp | -1.7 | 1.57 | 156.4 | 342.6 | 124.950 | 76.6 | 48.3 | 214.3 | -28.9 | 0.957 | -4.3 | -- |
| 20070727 | 2.7 | 2.8 | 0.80 | 8.3 | 269.9 | 123.717 | 23.8 | 37.8 | 303.7 | -9.3 | 0.566 | 5.1 | CAP |
| 20071021 | 1.2 | 2.5 | 0.80 | 172.9 | 94.6 | 27.535 | 63.4 | 37.9 | 95.6 | 20.1 | 0.519 | 4.6 | ORI |
| 20080325 | 3.7 | 1.9 | 0.48 | 7.3 | 358.8 | 184.634 | 7.7 | 36.3 | 95.8 | -13.2 | 0.997 | 2.9 | -- |
| 20080511 | hyp | -26.1 | 1.03 | 9.3 | 42.6 | 230.738 | 19.6 | 42.3 | 191.2 | -25.4 | 0.874 | -53.1 | -- |
| 20080812 | 1.0 | 3.7 | 0.74 | 110.8 | 157.7 | 139.878 | 56.4 | 38.9 | 41.3 | 57.8 | 0.981 | 6.4 | PER |
| 20081028 | 2.7 | 3.2 | 0.71 | 6.9 | 31.6 | 34.961 | 12.1 | 38.8 | 2.4 | -23.0 | 0.932 | 5.4 | -- |
| 20081102 | 3.5 | 1.8 | 0.82 | 5.2 | 118.2 | 40.086 | 27.9 | 36.1 | 52.7 | 14.6 | 0.335 | 3.3 | STA |
| 20081107 | hyp | -8.8 | 1.11 | 152.8 | 10.0 | 45.144 | 71.5 | 43.5 | 130.2 | 1.7 | 0.983 | -18.7 | -- |
| 20090428 | 3.1 | 2.4 | 0.68 | 10.6 | 244.7 | 37.916 | 18.0 | 37.3 | 211.2 | 8.9 | 0.773 | 4.1 | -- |
| 20090523 | hyp | -41.8 | 1.02 | 21.3 | 250.7 | 62.169 | 27.9 | 42.1 | 240.2 | 6.4 | 0.671 | -84.3 | -- |
| 20090812 | 0.5 | 5.4 | 0.82 | 111.9 | 154.0 | 139.617 | 57.5 | 39.8 | 44.0 | 57.8 | 0.967 | 9.8 | PER |
| 20090917 | 2.5 | 3.2 | 0.81 | 0.3 | 263.0 | 174.168 | 22.4 | 38.6 | 350.9 | -3.4 | 0.610 | 5.8 | -- |
| 20100421 | 0.4 | 93.7 | 0.99 | 71.7 | 257.8 | 30.847 | 45.3 | 41.9 | 252.1 | 21.2 | 0.610 | 186.9 | -- |
| 20100429 | 1.0 | 5.8 | 0.85 | 82.8 | 229.2 | 38.658 | 47.2 | 40.1 | 274.4 | 27.9 | 0.846 | 10.8 | -- |
| 20100530 | 2.5 | 3.0 | 0.86 | 4.6 | 284.4 | 68.652 | 28.2 | 38.1 | 254.4 | -18.4 | 0.429 | 5.6 | -- |
| 20110520 | 2.6 | 3.2 | 0.78 | 6.0 | 252.8 | 58.759 | 20.9 | 38.5 | 229.8 | -8.5 | 0.699 | 5.8 | -- |
| 20110630 | 1.5 | 7.9 | 0.87 | 45.0 | 187.8 | 97.886 | 28.6 | 40.4 | 274.9 | 61.5 | 1.012 | 14.7 | -- |
| 20110808 | 2.3 | 3.8 | 0.74 | 37.1 | 156.9 | 135.191 | 24.2 | 38.9 | 223.8 | 70.6 | 0.980 | 6.6 | -- |
| 20111005 | 2.6 | 2.8 | 0.85 | 7.9 | 103.2 | 11.464 | 27.2 | 38.2 | 19.9 | 0.1 | 0.437 | 5.2 | -- |
| 20111202 | 2.9 | 2.4 | 0.81 | 5.5 | 100.8 | 69.314 | 26.0 | 37.9 | 74.8 | 16.8 | 0.461 | 4.4 | -- |

Table S2: A summary of the event parameters. The table columns are as follows: (1 − 4) event date and time, (5 − 6) the shock source height and its uncertainty, (7) total range[*] and (8) horizontal range[*], (9) meteor azimuth, (10) meteor trajectory length, (11) meteor flight time, (12) observed infrasound signal travel time from the point of shock to the receiver, (13 − 14) observed infrasound signal back azimuth with its uncertainty, (15) measured infrasound signal dominant frequency, (16 − 17) measured infrasound signal dominant period with its uncertainty, (18 − 19) maximum signal amplitude with its uncertainty and (20 − 21) peak-to-peak amplitude with its uncertainty. The details outlining the methodology as to how these parameters were determined can be found in Silber and Brown [2014]. [*]Range denotes the distance in km from the point of shock along the meteor trajectory to the infrasound station.

| 1 | 2 | 3 | 4 | 5 | 6 | 7 | 8 | 9 | 10 | 11 | 12 | 13 | 14 | 15 | 16 | 17 | 18 | 19 | 20 | 21 |
|---|---|---|---|---|---|---|---|---|---|---|---|---|---|---|---|---|---|---|---|---|
| | | | | Shock Source Height | ± | Total Range* | Hor Range* | Meteor Azim | Meteor Traj Length | Meteor Flight Time | Sig Travel Time | Sig Back Azim | ± | Sig Freq | Sig Period | ± | Max Amp | ± | P2P Amp | ± |
| Date | Event Time | | | | | | | | | | | | | | | | | | | |
| | hh | mm | ss | km | km | km | km | ° | km | s | s | ° | ° | Hz | s | s | Pa | Pa | Pa | Pa |
| 20060419 | 7 | 5 | 56 | 54.4 | 1.1 | 83.7 | 63.5 | 98.9 | 35.6 | 3.30 | 280 | 144.5 | 0.0 | 9.6 | 0.12 | 0.07 | 0.10 | 0.04 | 0.13 | 0.07 |
| 20061104 | 3 | 29 | 29 | 77.0 | 1.1 | 116.4 | 87.2 | 293.8 | 17.1 | 1.03 | 354 | 293.1 | 0.1 | 6.4 | 0.16 | 0.00 | 0.06 | 0.02 | 0.09 | 0.03 |
| 20070125 | 10 | 2 | 5 | 102.7 | 0.5 | 126.6 | 73.7 | 340.0 | 109.5 | 1.64 | 398 | 289.4 | 4.1 | 0.9 | 1.35 | 0.03 | 0.06 | 0.01 | 0.08 | 0.02 |
| 20070727 | 4 | 51 | 58 | 85.0 | 1.5 | 149.0 | 122.5 | 350.8 | 31.3 | 1.57 | 467 | 29.9 | 0.4 | 1.1 | 0.94 | 0.08 | 0.06 | 0.01 | 0.09 | 0.03 |
| 20071021 | 10 | 26 | 25 | 101.2 | 1.4 | 194.8 | 166.3 | 24.5 | 19.8 | 0.70 | 700 | 38.9 | 3.1 | 0.5 | 2.21 | 0.03 | 0.11 | 0.00 | 0.20 | 0.01 |
| 20080325 | 0 | 42 | 3 | 61.6 | 0.6 | 113.1 | 94.9 | 15.3 | 43.3 | 5.21 | 341 | 305.0 | 4.0 | 7.0 | 0.14 | 0.01 | 0.10 | 0.06 | 0.14 | 0.12 |
| 20081028 | 3 | 17 | 35 | 52.7 | 3.6 | 73.3 | 50.9 | 0.4 | 60.9 | 5.21 | 240 | 306.5 | 0.7 | 11.1 | 0.10 | 0.01 | 0.06 | 0.06 | 0.10 | 0.12 |
| 20081102 | 6 | 13 | 26 | 85.0 | 0.5 | 193.1 | 173.4 | 356.7 | 17.9 | 1.33 | 576 | 292.2 | 9.5 | 2.3 | 0.44 | 0.03 | 0.08 | 0.04 | 0.12 | 0.08 |
| 20081107 | 7 | 34 | 16 | 81.9 | 0.6 | 119.9 | 87.6 | 297.2 | 58.5 | 0.97 | 378 | 330.4 | 0.6 | 1.5 | 0.59 | 0.05 | 0.06 | 0.01 | 0.09 | 0.01 |
| 20090523 | 7 | 7 | 25 | 78.1 | 2.3 | 134.7 | 109.8 | 40.2 | 21.1 | 1.14 | 428 | 62.2 | 0.1 | 3.0 | 0.35 | 0.02 | 0.20 | 0.04 | 0.32 | 0.07 |
| 20090917 | 1 | 20 | 38 | 76.6 | 2.1 | 132.6 | 108.2 | 296.9 | 28.5 | 1.60 | 429 | 358.0 | 2.1 | 3.0 | 0.37 | 0.12 | 0.08 | 0.02 | 0.13 | 0.04 |
| 20100421 | 4 | 49 | 43 | 86.3 | 0.8 | 216.6 | 198.6 | 281.5 | 37.0 | 1.07 | 709 | 5.7 | 0.5 | 1.0 | 1.21 | 0.27 | 0.05 | 0.01 | 0.07 | 0.01 |
| 20100429 | 5 | 21 | 35 | 93.0 | 1.9 | 191.9 | 167.9 | 268.4 | 18.0 | 0.50 | 617 | 320.4 | 0.5 | 0.9 | 0.99 | 0.35 | 0.07 | 0.01 | 0.11 | 0.02 |
| 20110520 | 6 | 2 | 9 | 94.5 | 0.7 | 180.0 | 153.2 | 21.6 | 14.3 | 0.80 | 565 | 62.5 | 0.3 | 1.9 | 0.65 | 0.02 | 0.03 | 0.01 | 0.06 | 0.01 |
| 20110630 | 3 | 39 | 38 | 87.7 | 0.5 | 161.7 | 135.8 | 209.2 | 12.4 | 0.87 | 535 | 186.2 | 0.3 | 3.0 | 0.40 | 0.00 | 0.03 | 0.01 | 0.04 | 0.02 |
| 20110808 | 5 | 22 | 6 | 63.6 | 0.3 | 179.1 | 167.4 | 156.1 | 53.3 | 2.84 | 565 | 170.2 | 0.4 | 1.5 | 0.50 | 0.03 | 0.03 | 0.01 | 0.04 | 0.02 |
| 20111005 | 5 | 8 | 53 | 77.8 | 4.2 | 131.3 | 105.7 | 342.2 | 29.5 | 1.50 | 407 | 306.6 | 0.8 | 4.6 | 0.18 | 0.01 | 0.13 | 0.06 | 0.20 | 0.12 |
| 20111202 | 0 | 31 | 4 | 64.0 | 0.6 | 148.3 | 133.7 | 265.1 | 101.8 | 4.40 | 449 | 339.1 | 0.8 | 2.8 | 0.38 | 0.11 | 0.15 | 0.06 | 0.21 | 0.12 |
| 20060805 | 8 | 38 | 50 | 87.8 | 0.6 | 124.7 | 88.3 | 268.5 | 28.5 | 0.80 | 427 | 255.5 | 1.3 | 0.9 | 1.19 | 0.10 | 0.49 | 0.10 | 0.68 | 0.20 |
| 20060805 | 8 | 38 | 50 | 101.4 | 0.4 | 130.0 | 81.2 | 268.5 | 28.5 | 0.80 | 450 | 257.0 | 0.2 | 0.6 | 2.01 | 0.17 | 0.10 | 0.02 | 0.18 | 0.05 |
| 20080511 | 4 | 22 | 17 | 94.6 | 0.4 | 111.7 | 59.3 | 20.6 | 41.5 | 1.97 | 371 | 24.8 | 0.4 | 1.3 | 0.77 | 0.03 | 0.03 | 0.01 | 0.04 | 0.01 |
| 20080511 | 4 | 22 | 17 | 88.9 | 0.5 | 114.7 | 72.5 | 20.6 | 41.5 | 1.97 | 381 | 24.8 | 0.4 | 2.9 | 0.36 | 0.00 | 0.01 | 0.00 | 0.02 | 0.01 |
| 20080812 | 8 | 19 | 29 | 86.2 | 0.8 | 157.0 | 131.3 | 224.9 | 12.4 | 0.47 | 554 | 249.4 | 0.1 | 1.6 | 0.59 | 0.05 | 0.02 | 0.00 | 0.03 | 0.01 |
| 20080812 | 8 | 19 | 29 | 87.9 | 0.8 | 157.3 | 130.4 | 224.9 | 12.4 | 0.47 | 557 | 249.8 | 0.5 | 1.8 | 0.58 | 0.01 | 0.02 | 0.00 | 0.03 | 0.01 |
| 20090428 | 4 | 43 | 37 | 60.7 | 6.2 | 138.6 | 124.8 | 351.0 | 29.0 | 2.97 | 456 | 53.6 | 1.7 | 4.1 | 0.21 | 0.03 | 0.16 | 0.03 | 0.22 | 0.06 |
| 20090428 | 4 | 43 | 37 | 70.9 | 1.1 | 140.6 | 121.5 | 351.0 | 29.0 | 2.97 | 460 | 55.3 | 1.5 | 2.7 | 0.32 | 0.01 | 0.06 | 0.02 | 0.11 | 0.04 |
| 20090812 | 7 | 55 | 58 | 80.6 | 0.3 | 155.9 | 133.4 | 225.7 | 17.7 | 0.57 | 522 | 204.5 | 1.2 | 1.6 | 0.59 | 0.05 | 0.09 | 0.02 | 0.16 | 0.03 |
| 20090812 | 7 | 55 | 58 | 80.5 | 0.3 | 155.9 | 133.4 | 225.7 | 17.7 | 0.57 | 525 | 204.3 | 1.1 | 1.8 | 0.58 | 0.01 | 0.07 | 0.01 | 0.09 | 0.02 |
| 20100530 | 7 | 0 | 31 | 92.7 | 2.4 | 156.4 | 125.9 | 16.3 | 32.5 | 1.33 | 530 | 324.4 | 0.4 | 1.5 | 0.64 | 0.01 | 0.03 | 0.01 | 0.06 | 0.02 |
| 20100530 | 7 | 0 | 31 | 92.1 | 2.8 | 156.6 | 126.7 | 16.3 | 32.5 | 1.33 | 535 | 325.2 | 0.1 | 1.9 | 0.84 | 0.21 | 0.03 | 0.01 | 0.05 | 0.01 |

Table S3: A summary of the photometric mass measurements using observed light curves from meteor video records. The columns are as follows (column numbers are in the first row): (1) meteor event date, (2) meteor velocity (km/s) at the onset of ablation, (3) begin height (km), (4) end height (km), (5) instrument corrected peak magnitude ($M_{HAD}$) from the observed light curve, (6-7) minimum and maximum values of shape coefficient ($K$), and (8-9) minimum and maximum values of ablation coefficient (σ) used in the FM model fitting, (10) estimated meteor mass (g) from equation (8), (11) estimated meteor mass (g) using integrated light curve, (12) initial (entry) meteor mass (g) from the FM model, (13) PE coefficient and (14) meteoroid group type.

| 1 | 2 | 3 | 4 | 5 | 6 | 7 | 8 | 9 | 10 | 11 | 12 | 13 | 14 |
|---|---|---|---|---|---|---|---|---|---|---|---|---|---|
| | Entry | H | H | | | | | | | | FM | | |
| | Velocity | begin | end | Peak | | | | | Mass | Mass | Model | | |
| Date | (km/s) | (km) | (km) | Mag | K min | K max | σ min | σ max | (JVB) (g) | (int) (g) | Mass (g) | PE | Type |
| 20060419 | 14.2 | 72.0 | 47.7 | -2.7 | 0.14 | 0.53 | 0.009 | 0.009 | 107.4 | 23.5 | 20.0 | -4.69 | II |
| 20060805 | 67.5 | 126.4 | 74.5 | -12.8 | 1.4 | 2.3 | 0.002 | 0.004 | 5927.6 | 432.9 | 74.0 | -6.20 | IIIB |
| 20061104 | 30.3 | 89.9 | 65.8 | -7.2 | 1.8 | 4.18 | 0.002 | 0.005 | 459.9 | 12.5 | 12.0 | -6.39 | IIIB |
| 20070125 | 71.2 | 119.2 | 88.5 | -5.9 | 1.99 | 3.69 | 0.001 | 0.004 | 9.5 | 2.7 | 0.9 | -5.12 | II |
| 20070727 | 26.3 | 96.2 | 70.6 | -8.2 | 0.46 | 2.68 | 0.002 | 0.004 | 2583.9 | 91.5 | 63.0 | -6.22 | IIIB |
| 20071021 | 64.3 | 130.8 | 81.7 | -8.8 | 2.5 | 2.5 | 0.004 | 0.004 | 57.5 | 10.6 | 4.3 | -5.82 | IIIB |
| 20080325 | 13.5 | 76.2 | 32.8 | -5.9 | 0.69 | 0.79 | 0.015 | 0.015 | 2912.0 | 792.9 | 917.0 | -4.51 | I |
| 20080511 | 23.5 | 95.2 | 77.3 | -3.8 | 2.54 | 2.54 | 0.06 | 0.06 | 85.8 | 5.2 | 8.0 | -5.90 | II |
| 20080812 | 56.6 | 105.7 | 82.0 | -1.8 | 3.29 | -- | 0.009 | -- | 0.2 | 0.1 | 0.1 | -4.93 | II |
| 20081028 | 15.4 | 81.2 | 41.1 | -4.1 | 0.66 | 0.66 | 0.014 | 0.014 | 309.8 | 79.6 | 110.0 | -4.40 | I |
| 20081102 | 30.1 | 96.5 | 62.6 | -7.7 | 1.73 | 2.05 | 0.002 | 0.002 | 663.9 | 53.3 | 18.0 | -5.58 | IIIA |
| 20081107 | 71.6 | 113.5 | 81.5 | -3.1 | 1.99 | 2.59 | 0.003 | 0.006 | 0.4 | 0.2 | 0.1 | -4.56 | I |
| 20090428 | 21.2 | 83.5 | 38.0 | -7.2 | 0.14 | 1.09 | 0.003 | 0.005 | 3086.5 | 784.1 | 330.0 | -4.77 | II |
| 20090523 | 29.9 | 95.9 | 72.4 | -2.0 | 0.2 | 0.86 | 0.042 | 0.044 | 2.7 | 0.7 | 2.2 | -5.10 | II |
| 20090812 | 58.7 | 108.5 | 80.4 | -6.7 | 1.29 | 3.29 | 0.008 | 0.01 | 20.6 | 3.4 | 1.8 | -5.65 | IIIA |
| 20090917 | 24.2 | 85.7 | 72.4 | -2.7 | 3.05 | 3.05 | 0.004 | 0.004 | 20.7 | 6.6 | 8.5 | -5.31 | IIIA |
| 20100421 | 45.9 | 108.5 | 74.6 | -9.3 | 1.5 | 1.5 | 0.005 | 0.0055 | 861.5 | 45.7 | 17.0 | -5.95 | IIIB |
| 20100429 | 47.7 | 105.7 | 89.9 | -2.6 | 4.89 | 4.89 | 0.014 | 0.014 | 0.9 | 0.2 | 0.3 | -5.79 | IIIB |
| 20100530 | 29.3 | 96.0 | 78.3 | -0.8 | 2.19 | 2.99 | 0.01 | 0.012 | 1.2 | 0.3 | 0.3 | -5.12 | II |
| 20110520 | 22.5 | 95.7 | 84.1 | -3.1 | 2.79 | 2.99 | 0.039 | 0.039 | 21.3 | 2.3 | 2.5 | -6.35 | IIIB |
| 20110630 | 29.8 | 100.5 | 71.7 | -7.8 | 2.49 | 2.49 | 0.003 | 0.003 | 527.5 | 18.0 | 10.0 | -6.05 | IIIB |
| 20110808 | 25.5 | 86.6 | 39.9 | -9.3 | 0.99 | 1.49 | 0.002 | 0.002 | 9990.9 | 2586.4 | 1003.0 | -4.77 | II |
| 20111005 | 28.5 | 96.2 | 64.5 | -2.9 | 1.09 | 1.59 | 0.004 | 0.004 | 6.8 | 2.6 | 20.0 | -4.78 | II |
| 20111202 | 27.6 | 97.0 | 53.8 | -3.1 | 0.29 | 0.69 | 0.008 | 0.009 | 18.0 | 8.8 | 9.0 | -4.06 | I |

Table S4: The range of PE parameter and the associated meteoroid group presumed type, type of the material and representative density as given by Ceplecha et al. [1998]. The percentage of observed meteoroids in each group according to mass category typical of fireball networks (mass range 0.1 - 2000 kg) and small cameras (mass range $10^{-4}$ kg to 0.5 kg) as published in [Ceplecha et al., 1998] and the percentages found in this study (masses $10^{-4}$ kg to 1.0 kg). The range of values for the intrinsic ablation coefficients ($\sigma$) and the shape density coefficient ($K$) found in our study within the framework of the FM model are given in the last two columns.

| PE range | Group type | Type of the meteoroid material | $\rho_m$ (kg/m³) | % observed (fireball networks) | % observed (small cameras) | % observed (this study) | $\sigma$ | K |
|---|---|---|---|---|---|---|---|---|
| PE > -4.6 | I | ordinary chondrites, asteroids | 3700 | 29 | 5 | 13 | 0.006-0.021 | 0.46- 1.29 |
| -4.6 ≥ PE > -5.25 | II | carbonaceous chondrites, comets, asteroids | 2000 | 33 | 39 | 37 | 0.002-0.19 | 0.1- 3.09 |
| -5.25 ≥ PE > -5.7 | IIIa | regular cometary material | 750 | 26 | 41 | 13 | 0.002-0.009 | 1.93- 3.29 |
| PE ≤ -5.7 | IIIb | soft cometary material | 270 | 9 | 19 | 37 | 0.001-0.06 | 1.2- 4.89 |

Table S5: A summary of $R_0$ as derived from the FM model. The associated outputs from top-down modelling are also shown. The two entries in the Taltitude, which is the transition height, in the "New d'" section denote events that never transition to the linear regime. Therefore, the original definition for d' was used by default. [*]There was no convergent solution.

| Date | Ro | Ro err | Original d' [Towne, 1967] | | | | | New d' | | | | |
|------|-----|--------|------|------|-------|-------|-----------|------|------|-------|-------|-----------|
| | | | Tlin | Tws | Alin | Aws | Taltitude | Tlin | Tws | Alin | Aws | Taltitude |
| 20060419 | 1.09 | 0.05 | 0.09 | 0.09 | 0.234 | 0.119 | 10.0 | 0.09 | 0.09 | 0.125 | 0.119 | 1.1 |
| 20061104 | 2.45 | 0.01 | 0.17 | 0.19 | 0.216 | 0.031 | 25.4 | 0.18 | 0.19 | 0.048 | 0.031 | 7.1 |
| 20070125 | 25.95 | 0.36 | 1.05 | 1.26 | 1.363 | 0.021 | 55.0 | 1.22 | 1.26 | 0.053 | 0.021 | 12.8 |
| 20070727 | 11.48 | 0.21 | 0.68 | 0.80 | 0.731 | 0.04 | 38.4 | 0.79 | 0.80 | 0.054 | 0.04 | 5.0 |
| 20071021 | 66.66 | 5.36 | 2.24 | 2.65 | 2.202 | 0.036 | 54.5 | 2.53 | 2.65 | 0.148 | 0.036 | 19.3 |
| 20080325 | 5.79 | 0.39 | 0.32 | 0.35 | 0.639 | 0.195 | 16.1 | 0.33 | 0.35 | 0.452 | 0.195 | 12.0 |
| 20081028 | 2.52 | 0.22 | 0.15 | 0.16 | 0.582 | 0.272 | 10.9 | 0.16 | 0.16 | 0.317 | 0.272 | 2.6 |
| 20081102 | 5.30 | 0.55 | 0.36 | 0.40 | 0.23 | 0.017 | 33.2 | 0.38 | 0.40 | 0.091 | 0.017 | 22.2 |
| 20081107 | 0.56 | 0.43 | 0.06 | 0.06 | 0.044 | 0.006 | 25.8 | 0.06 | 0.06 | 0.007 | 0.006 | 1.3 |
| 20090523 | 1.99 | 0.36 | 0.17 | 0.19 | 0.167 | 0.022 | 26.5 | 0.18 | 0.19 | 0.094 | 0.022 | 19.5 |
| 20090917 | 4.68 | 0.47 | 0.33 | 0.36 | 0.39 | 0.047 | 27.8 | 0.35 | 0.36 | 0.105 | 0.047 | 11.9 |
| 20100421 | 15.88 | 0.07 | 0.85 | 0.96 | 0.587 | 0.036 | 36.2 | 0.95 | 0.96 | 0.044 | 0.036 | 3.5 |
| 20100429 | 8.87 | 0.79 | 0.53 | 0.61 | 0.387 | 0.015 | 42.5 | 0.59 | 0.61 | 0.037 | 0.015 | 12.7 |
| 20110520 | 9.97 | 0.20 | 0.56 | 0.65 | 0.439 | 0.015 | 42.0 | 0.63 | 0.65 | 0.038 | 0.015 | 11.9 |
| 20110630 | 7.42 | 0.05 | 0.46 | 0.53 | 0.392 | 0.018 | 40.6 | 0.53 | 0.53 | 0.021 | 0.018 | 2.4 |
| 20110808 | 12.32 | 0.06 | 0.62 | 0.68 | 0.9 | 0.254 | 17.2 | 0.62 | 0.68 | 0.9 | 0.254 | 17.2[*] |
| 20111005 | 7.69 | 0.82 | 0.42 | 0.46 | 0.565 | 0.065 | 28.3 | 0.44 | 0.46 | 0.285 | 0.065 | 20.0 |
| 20111202 | 2.24 | 0.04 | 0.16 | 0.17 | 0.19 | 0.062 | 15.4 | 0.16 | 0.17 | 0.075 | 0.062 | 3.4 |
| 20060805 | 12.50 | 0.44 | 0.64 | 0.75 | 0.844 | 0.038 | 41.2 | 0.69 | 0.75 | 0.228 | 0.038 | 23.9 |
| 20060805 | 59.42 | 0.76 | 1.88 | 2.29 | 2.861 | 0.044 | 56.5 | 2.20 | 2.29 | 0.132 | 0.044 | 15.4 |
| 20080511 | 13.10 | 0.03 | 0.63 | 0.74 | 0.932 | 0.028 | 46.9 | 0.73 | 0.74 | 0.034 | 0.028 | 3.7 |
| 20080511 | 11.90 | 0.10 | 0.61 | 0.70 | 0.895 | 0.037 | 42.0 | 0.61 | 0.70 | 0.895 | 0.037 | 42.0[*] |
| 20080812 | 5.75 | 0.19 | 0.38 | 0.43 | 0.326 | 0.02 | 37.0 | 0.43 | 0.43 | 0.021 | 0.02 | 1.4 |
| 20080812 | 6.34 | 0.15 | 0.40 | 0.46 | 0.348 | 0.018 | 39.1 | 0.46 | 0.46 | 0.02 | 0.018 | 2.2 |
| 20090428 | 5.11 | 1.08 | 0.31 | 0.33 | 0.486 | 0.171 | 14.6 | 0.32 | 0.33 | 0.306 | 0.171 | 8.7 |
| 20090428 | 2.28 | 0.01 | 0.18 | 0.19 | 0.165 | 0.037 | 20.2 | 0.19 | 0.19 | 0.043 | 0.037 | 2.8 |
| 20090812 | 0.76 | 0.08 | 0.08 | 0.09 | 0.05 | 0.007 | 26.7 | 0.09 | 0.09 | 0.008 | 0.007 | 2.6 |
| 20090812 | 0.75 | 0.07 | 0.08 | 0.09 | 0.049 | 0.007 | 26.6 | 0.09 | 0.09 | 0.011 | 0.007 | 7.6 |
| 20100530 | 3.35 | 0.13 | 0.26 | 0.31 | 0.198 | 0.008 | 41.8 | 0.30 | 0.31 | 0.014 | 0.008 | 7.7 |
| 20100530 | 3.29 | 0.14 | 0.26 | 0.30 | 0.196 | 0.009 | 41.1 | 0.30 | 0.30 | 0.011 | 0.009 | 3.6 |